\documentclass[aps,superscriptaddress,reprint,
               colorlinks=true,
               linkcolor=black,
               citecolor=blue,
               urlcolor=black]{revtex4-2}

\usepackage{graphicx}
\usepackage{dcolumn}
\usepackage{bm}
\usepackage{amsmath,amssymb}
\usepackage{xcolor}
\usepackage{pifont}
\usepackage[bottom]{footmisc}
\usepackage{booktabs}
\usepackage{multirow}
\usepackage{array}
\usepackage{caption}
\DeclareCaptionLabelSeparator{pipe}{\,\textbar\,}
\usepackage{float}
\usepackage[rightcaption]{sidecap}
\sidecaptionvpos{figure}{t}
\usepackage{titlesec}
\titleformat{\section}
  {\normalfont\bfseries\normalsize}
  {}{0em}{}
\titlespacing*{\section}
  {0pt}{10pt plus 2pt minus 2pt}{2pt}
\titleformat{\subsection}
  {\normalfont\bfseries\normalsize}
  {}{0em}{}
\titlespacing*{\subsection}
  {0pt}{4pt plus 1pt minus 1pt}{1pt}

\setcitestyle{super,square}      


\preprint{APS/123-QED}
\def\CVS{CsV$_3$Sb$_5$}
\def\KVS{KV$_3$Sb$_5$}
\def\RVS{RbV$_3$Sb$_5$}
\def\KRC{K$_{0.20}$Rb$_{0.43}$Cs$_{0.37}$V$_3$Sb$_5$}
\def\AVS{$A$V$_3$Sb$_5$}
\def\cm{cm$^{-1}$}
\def\Tc{$T_{\rm CDW}$}
\def\DCDW{$\Delta_{\rm CDW}$}

\usepackage{graphicx}%
\usepackage{nicefrac}
\usepackage{lmodern}
\usepackage{amssymb}
\usepackage[T1]{fontenc}
\usepackage{amsmath}
\usepackage{amsfonts}
\usepackage[note-name=, use-sort-key = false]{notes2bib}
\usepackage{xcolor}
\usepackage{color}
\usepackage[colorlinks,bookmarks=false,citecolor=darkblue,linkcolor=red,urlcolor=blue]{hyperref} 
\definecolor{darkred}{rgb}{0.7,0.0,0.0}

\definecolor{darkblue}{rgb}{0,0.02,0.45}

\definecolor{darkgreen}{rgb}{0.02,0.45,0.0}

\definecolor{violet}{rgb}{0.8,0.2,0.6}

\begin{document}

\title{Band-Saddle-Point Engineering in Mixed $A$-site $A$V$_3$Sb$_5$ Kagome Metals}

\author{Maxim Wenzel}
\email{maxim.wenzel@pi1.physik.uni-stuttgart.de}
\affiliation{1. Physikalisches Institut, Universität Stuttgart, D-70569 Stuttgart, Germany}

\author{Alexander A. Tsirlin}
\affiliation{Felix Bloch Institute for Solid-State Physics, Leipzig University, 04103 Leipzig, Germany}

\author{Andrea N. Capa Salinas}
\affiliation{ Materials Department, University of California Santa Barbara, Santa Barbara, California 93106, USA}

\author{Brenden R. Ortiz}
\affiliation{Materials Science and Technology Division, Oak Ridge National Laboratory, Oak Ridge, Tennessee 37831, USA}

\author{Stephen D. Wilson}
\affiliation{ Materials Department, University of California Santa Barbara, Santa Barbara, California 93106, USA}

\author{Ece Uykur}
\affiliation{Helmholtz-Zentrum Dresden-Rossendorf, Institute of Ion Beam Physics and Materials Research, 01328 Dresden, Germany}

\author{Martin Dressel}
\affiliation{1. Physikalisches Institut, Universität Stuttgart, D-70569 Stuttgart, Germany}

\date{\today}

\begin{abstract}
\AVS\ compounds ($A =$~K, Rb, Cs) offer an experimental platform for probing the physics of kagome metals near ideal Van Hove filling. Here, we use optical spectroscopy on \AVS\ with mixed $A$-site compositions to reveal the effect of the alkali metal on the low-energy electronic structure. Supported by density-functional-theory calculations, we identify a band-saddle-point inversion as a chemical consequence of Cs substitution, revealing that $A$-site composition controls the orbital character of Van Hove singularities, Fermi surface topology, and electron-phonon interactions while leaving the kagome layer intact.
\end{abstract}

\maketitle

\section{INTRODUCTION}
Vanadium-based 135 kagome metals feature a quasi-two-dimensional band structure owing to the large separation between neighboring kagome layers and the absence of magnetic ordering~\cite{Wilson2024, DiSante2026, Jiang2022}. Furthermore, the proximity of the Fermi level to Van Hove singularities leads to various exotic phases predicted for kagome metals theoretically~\cite{Zhan2026, Profe2024, Kiesel2013, Denner2021, Park2021}. Experimentally, charge-density-wave (CDW) order accompanied by spontaneous rotational symmetry breaking (nematicity)~\cite{Nie2022, Li2022, Li2023, Wulferding2022, Xu2022}, chiral CDW states with a possible breaking of time-reversal symmetry~\cite{Jiang2021, Mielke2022, Xu2022, Shumiya2021}, and unconventional multigap superconductivity at low temperatures~\cite{Wilson2024, Jiang2022} have been observed.

The CDW state in \AVS\ is accompanied by a 2~$\times$~2 in-plane lattice modulation, predominantly of the vanadium sublattice, with a star-of-David (soD) or tri-hexagonal (trh) pattern, as suggested by ab-initio calculations~\cite{Ortiz2021, Tan2021, Kang2023}. Additionally, an out-of-plane modulation is observed, resulting in a three-dimensional CDW state with a staggered tri-hexagonal arrangement for \KVS\ and \RVS~\cite{Kang2023, Kautzsch2023, Frassineti2023}. In contrast, \CVS\ exhibits considerably more intricate CDW phases that depend on stacking disorder and thermal history with coexisting trh and soD layers~\cite{Ortiz2021, Kang2023, Luo2022, Kautzsch2023, Stahl2022, Xiao2023, Hu2022, Plumb2024}. Moreover, the CDW state of \CVS\ is accompanied by multiple unique features such as an anomalous Nernst effect~\cite{Zhou2022, Chen2022, Gan2021}, anomalies in NMR spectra~\cite{Song2022}, and a unidirectional charge order observed in surface-sensitive studies~\cite{Li2023, Li2022a, Zhao2021, Hu2022a, Guo2025}.

In the normal state, \CVS\ displays lower resistivity and significantly reduced electronic correlations in comparison to its K and Rb congeners~\cite{Wenzel2022, Uykur2021, Uykur2022, Wenzel2023}. Most notably, optical spectroscopy revealed a pronounced change in the interband transitions in \CVS\ compared to \KVS\ and \RVS~\cite{Wenzel2022, Uykur2021, Uykur2022}. Together, these observations highlight distinct electronic properties of \CVS, and motivate a closer examination of the interplay between chemical composition, lattice and electronic degrees of freedom in governing the evolution of the low-energy electronic structure across the \AVS\ family.

\begin{figure*}
\centering 
\includegraphics{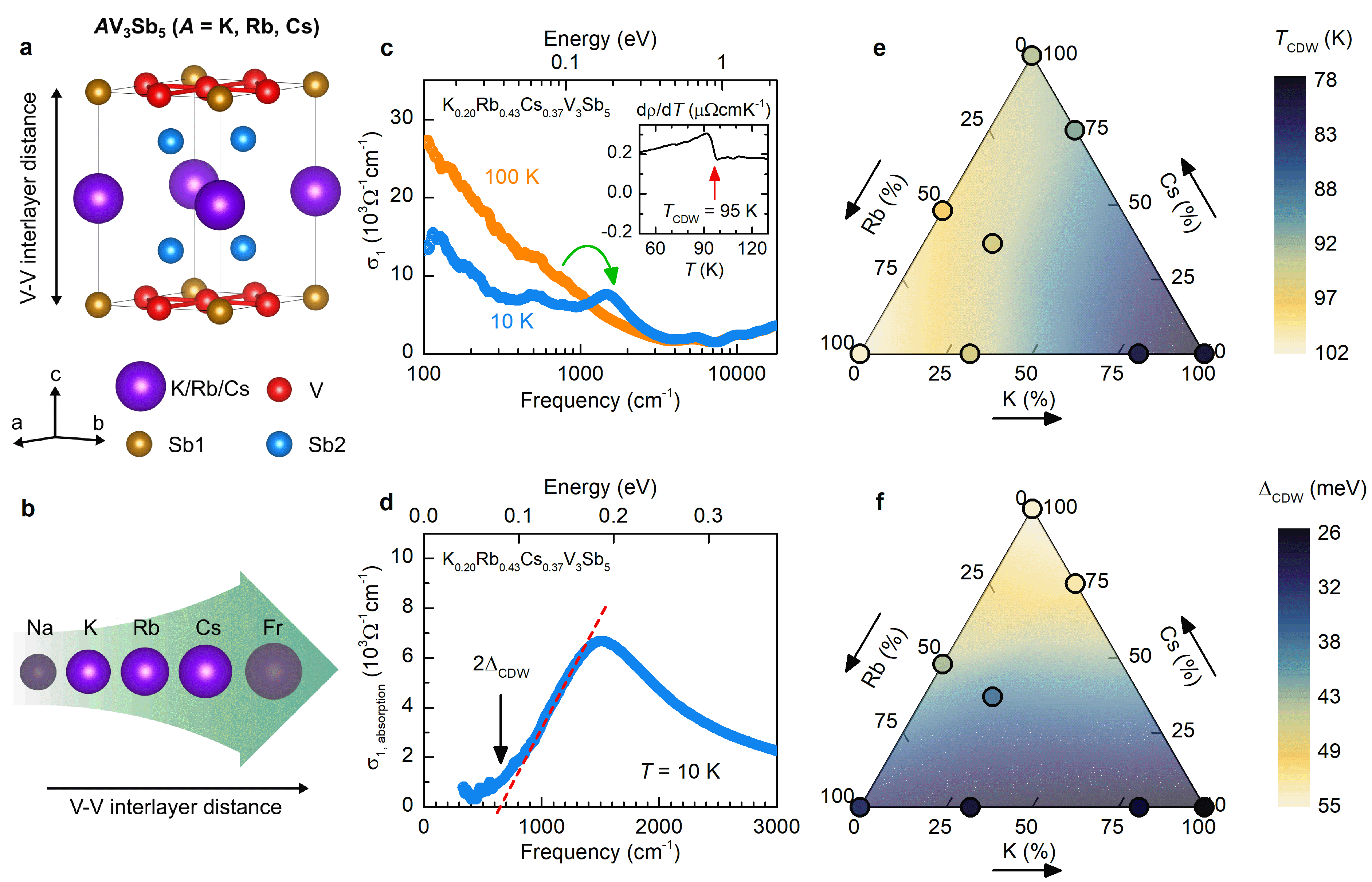}
\caption{\textbf{Evolution of the CDW state in mixed $A$-site \AVS\ systems.} \textbf{a}~Crystal structure of \AVS\ compounds highlighting the V kagome layers (red) separated by alkali ions (purple) and Sb2 atoms (blue), visualized by \texttt{VESTA}~\cite{Momma2008}. \textbf{b}~Evolution of atomic radius of alkali ions and consequences for the V-V interlayer distance. \textbf{c}~Experimental optical conductivity of \KRC\ at 10~K ($T$ < \Tc), and above the CDW transition (100~K). The green arrow indicates the spectral weight transfer due to the gap opening. \textbf{d}~Absorption edge after subtracting the low-energy contributions from the spectrum at 10~K. The extrapolated zero crossing (red-dashed line) corresponds to 2$\Delta_{\mathrm{CDW}}$. \textbf{e}~Contour plot showing the evolution of \Tc\ as a function of alkali ion concentration. Colored dots represent experimentally determined values from the dc resistivity curves (see Supplementary Table~S2) as outlined in the inset of panel \textbf{c} and Fig.~S2. \textbf{f}~Alkali-site-content-evolution of the CDW gap energy determined from the absorption edge presented in panel \textbf{d}. Experimental values (colored dots) are obtained at 10~K and are listed in Supplementary Table~S2. Values for the parent compounds are taken from previous optical studies~\cite{Uykur2021, Uykur2022, Wenzel2022}.}
\label{Fig1}
\end{figure*}

Mixed $A$-site compounds, in which different alkali ions occupy the $A$ site, provide a unique opportunity to investigate how alkali-ion substitution influences the low-energy electronic structure and CDW formation in \AVS\ compounds. They significantly expand the accessible chemical space and, more generally, offer a clean tuning parameter that primarily modifies the kagome interlayer distance, as illustrated in Fig.~\ref{Fig1}\textbf{a} and \textbf{b}, while only minimally perturbing the kagome network itself~\cite{Ortiz2023}. Here, we investigate the influence of $A$-site composition on the low-energy electrodynamics using infrared spectroscopy complemented by density-functional-theory (DFT) calculations.

Our results reveal a continuous evolution of physical properties across the series and identify the inversion of the band saddle points as a chemical consequence of Cs substitution, despite the absence of alkali-metal states near the Fermi level and only marginal changes in lattice parameters. The resulting band structure of Cs-containing compounds gives rise to unique low-energy interband transitions, an enhanced CDW gap, and significant changes in the Fermi-surface topology. We further demonstrate the strong sensitivity of the Fermi surface and electron–phonon coupling to $A$-site composition, providing insights into the interplay between electronic structure, lattice degrees of freedom, and the CDW state in \AVS\ compounds.

\section{RESULTS}
\subsection{Charge-Density-Wave Gap}
Five mixed A-site samples with varying alkali-ion concentration were characterized
by dc resistivity measurements to determine the CDW transition temperature. The transition is marked by a kink in the resistivity, most clearly resolved in its first derivative, as shown for \KRC\ in the inset of Fig.~\ref{Fig1}\textbf{c} as an example. The extracted \Tc\ evolves smoothly with alkali-site composition~\cite{Ortiz2023}, connecting the three parent compounds as illustrated in Fig.~\ref{Fig1}\textbf{e}. In addition, increasing Cs content systematically reduces the electrical resistivity, indicating enhanced metallicity (see Supplementary Fig.~S2).

The optical spectra of \AVS\ compounds exhibit pronounced changes across the
CDW transition. Below \Tc, a textbook-like gap opening is observed as spectral weight is transferred from low to high energies, giving rise to a new absorption peak around 1500~\cm, as illustrated in Fig.~\ref{Fig1}\textbf{c} for \KRC. The CDW gap 2\DCDW\ is extracted from the onset of this absorption after subtracting unrelated low-energy contributions using the Drude-Lorentz analysis (see Supplementary Note~IIA), as outlined in Fig.~\ref{Fig1}\textbf{d}. The extracted gap  evolves continuously but nonlinearly with alkali-site composition, illustrated in Fig.~\ref{Fig1}\textbf{f}. While it remains nearly unchanged between Cs-free compounds, it increases markedly with the addition of Cs, reaching almost twice the value in pristine \CVS. This highlights the similarity of the CDW state in the K- and Rb-based compounds and its substantial modification upon Cs substitution. While the absolute value of the gap energy depends on the method of extraction, the overall evolution across alkali-site compositions remains robust, as discussed in Supplementary Note~II.

\begin{figure*}
\centering 
\includegraphics{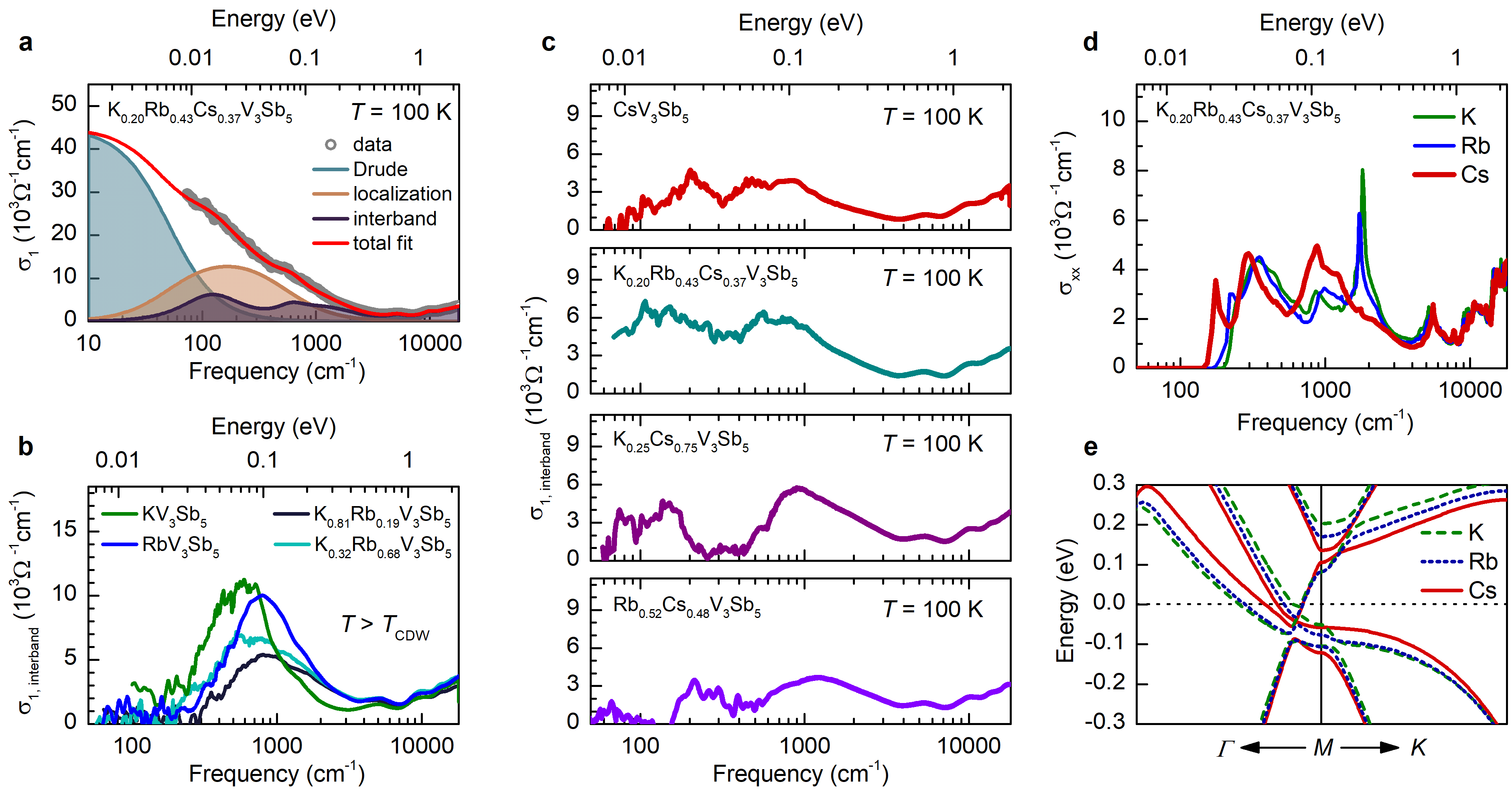}
\caption{\textbf{Evolution of interband transitions in the normal state.} \textbf{a}~Decomposed optical conductivity of \KRC\ at 100~K modeled by a Drude + localization peak (intraband) and several Lorentzians describing the interband transitions. \textbf{b}~Experimental interband optical conductivity of Cs-free compounds obtained by subtracting the intraband contributions from the data. All data are taken above the CDW transition. \textbf{c}~Same as panel \textbf{b} for Cs-containing compounds. The data of the parent compounds are taken from our previous optical studies~\cite{Uykur2021, Uykur2022, Wenzel2022}. \textbf{d}~Interband optical conductivity calculated by DFT using the lattice parameters of \KRC. Different alkali ions were used in the calculations as described in the Methods section. \textbf{e}~Corresponding band structures focusing on the saddle point region.}
\label{Fig2}
\end{figure*} 

\subsection{Normal-State Interband Transitions}
Having discussed the evolution of the CDW state with varying $A$-site composition, we now examine modifications of the normal-state electronic band structure through the interband optical conductivity. The interband contribution is obtained by subtracting the modeled Drude and localization peaks (intraband contribution) from the spectra (see Fig.~\ref{Fig2}\textbf{a} and Supplementary Note~IIA). The resulting interband transitions for Cs-free compounds are shown in Fig.~\ref{Fig2}\textbf{b}, while those of the Cs-containing compounds are presented in Fig.~\ref{Fig2}\textbf{c}. Note that the analysis is performed at the lowest temperature just above \Tc, where reduced scattering sharpens the intraband response and facilitates the identification of interband features.

At high energies, the optical spectra are nearly indistinguishable down to approximately 1000~\cm, where all compounds exhibit a broad absorption peak. At lower energies, however, distinct interband features emerge around 200~\cm\ only in Cs-containing compounds. In pristine \CVS, these absorptions have been associated with the inverted vanadium band saddle points~\cite{Wenzel2022, Uykur2021}. The observation of analogous low-energy absorptions in the mixed compounds therefore suggests that the electronic structure associated with the inverted saddle points may persist in all Cs-containing compounds.

To disentangle structural and chemical effects, we performed DFT calculations of the electronic structure using fixed experimental lattice parameters of the mixed $A$-site compounds, while varying the alkali-site ions. Fig.~\ref{Fig2}\textbf{d} shows representative results for the lattice parameters of \KRC. Note that calculations based on the other experimental crystal structures lead to the same qualitative results (see Supplementary Note~IIIB). In agreement with the experimental observation, only calculations including Cs reproduce the low-energy interband transitions in Cs-containing compositions. Replacing Cs by K or Rb in the calculation instead yields a sharp absorption peak near 2000~\cm, inconsistent with the experiments. Furthermore it suppresses the spectral weight around 1000~\cm, and shifts the onset of interband transitions to higher energies, thus removing the spectral weight from the 100-200~\cm range where it is clearly observed experimentally.

\begin{figure*}
\centering 
\includegraphics{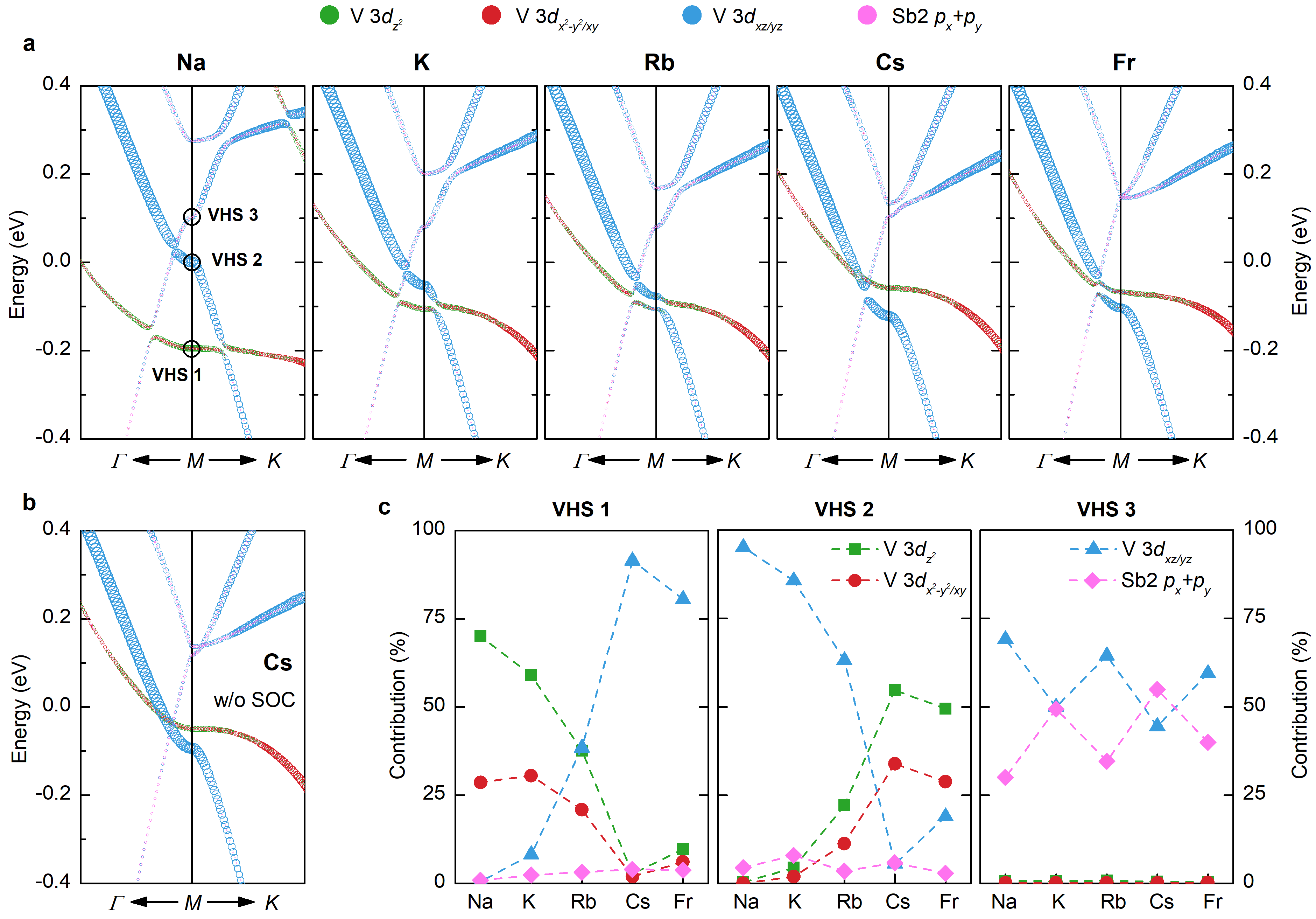}
\caption{\textbf{Evolution of the saddle-point electronic structure.} \textbf{a}~Low-energy band structures of \AVS\ compounds in the saddle-point region with colored dots representing contributions from different atomic orbitals. Experimental lattice parameters of \KRC\ were used for all calculations, while the alkali ion was varied from Na to Fr (left to right). \textbf{b}~Calculated band dispersion in the case of Cs in the absence of spin-orbit coupling. \textbf{c}~Evolution of the orbital character at the saddle points as a function of the alkali-site element. }
\label{Fig3}
\end{figure*}

While the inclusion of Cs reproduces the experimentally observed low-energy interband transitions, quantitative discrepancies remain in both intensity and energy. In particular, at very low energies, the experimentally observed interband transitions appear at slightly lower energies than predicted by DFT. These observations are consistent with previous optical studies evidencing that \CVS\ is well described by band theory, whereas \KVS\ and \RVS\ show prominent deviations that can be nevertheless captured by a rigid upward shift of the Fermi level, mimicking the effects of orbital-selective band-energy renormalization~\cite{Uykur2021, Uykur2022, Wenzel2022}. Applying the same approach to the mixed K/Rb compounds likewise improves the agreement with experiment (see Supplementary Fig.~S15). In contrast, for Cs-containing compounds, a slight rescaling of the energy axis, shifting the calculated interband transitions to lower energies, provides better overall agreement, as discussed in Supplementary Note~IIIB. Importantly, these modifications do not influence the overall band dispersion or the orbital character of the electronic states and hence do not affect the qualitative difference between Cs-containing and Cs-free compounds with DFT calculations.

\subsection{$A$-Site Controlled Band Inversion}
Further insight into the role of the alkali ion in determining the electronic band structure is provided by orbital-resolved band structures, presented in Fig.~\ref{Fig3}\textbf{a}, which display the atomic orbital contributions to the electronic states. Here, we focus on the $M$-point region, where three saddle points, corresponding to Van Hove singularities (VHS), are observed, two below and one above the Fermi level. We label these VHSs in order of increasing energy, i.e., VHS1, VHS2, and VHS3. To systematically track their evolution with alkali-ion size, we performed calculations for the experimentally relevant K, Rb, and Cs ions, as well as for the hypothetical Na and Fr analogues. While VHS3 remains of mixed V 3$d_{xz/yz}$ (blue) and Sb2 $p_x+p_y$ (pink) character, pronounced changes occur for VHS1 and VHS2. In particular, VHS2 evolves from predominantly V $3d_{xz/yz}$ (blue) character to $3d_{z^2}$ (green) and $3d_{x^2-y^2/xy}$ (red) character with increasing alkali-ion radius; correspondingly, the predominantly V $3d_{z^2}$ (green) and $3d_{x^2-y^2/xy}$ (red) character associated with VHS1 for Na and K shifts to VHS2 for Cs and Fr (Fig.~\ref{Fig3}\textbf{c}). Interestingly, Rb represents an intermediate case, where all three saddle points exhibit strongly mixed orbital character, indicating a gradual inversion of the saddle points rather than an abrupt change. 

Overall, alkali substitution primarily modifies the orbital character of the saddle points while leaving their energies largely unaffected (with Na being the exception). This behavior closely resembles DFT results for the parent compounds (using their experimental lattice parameters)~\cite{Wenzel2022, Sim2024}, suggesting that the evolution of the saddle points is primarily associated with the alkali-ion site rather than with changes in lattice parameters. Relativistic effects are enhanced in heavier alkaline metals, but it does not explain the observed behavior. Indeed, band inversion persists in calculations without spin-orbit coupling (SOC), with SOC only gapping out the knot of bands along $M \rightarrow \Gamma$ as shown in Fig.~\ref{Fig3}\textbf{b}.

The modifications of the band saddle points can be understood as an upward shift of the 'green' V 3$d_{z^2}$ band, evolving continuously with increasing ionic radius of the alkaline metal. As discussed in Supplemental Note~IIIA, this shift is not restricted to the low-energy $M$-point region, but can be traced along several directions in $k$-space near $K$, $L$, and $H$. Here, however, the relevant bands lie at higher energies and therefore do not directly affect the Fermi surface.

\subsection{Fermi Surface and Electron-Phonon Coupling}
\begin{figure*}
\centering 
\includegraphics{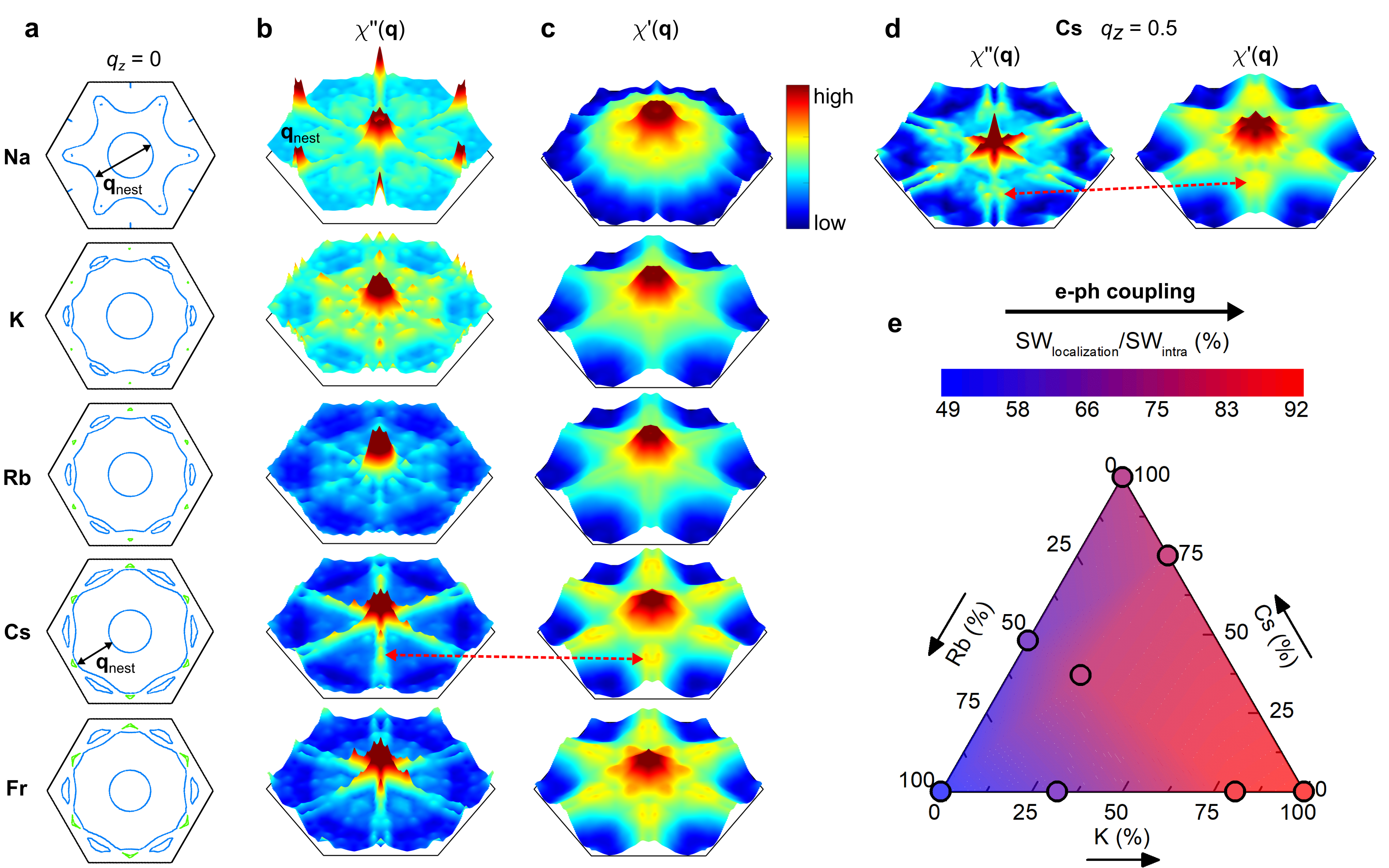}
\caption{\textbf{Fermi surface nesting and electron-phonon coupling.} \textbf{a}~Fermi surfaces for different alkali ions at $q_z = 0$, illustrated using \texttt{FermiSurfer}~\cite{Kawamura2019}. Well-defined nesting vectors observed in the case of using Na and Cs for the calculations are marked by black arrows. \textbf{b}~Imaginary part of the electronic susceptibility as a function of $q_x$ and $q_y$, with $q_z = 0$. \textbf{c}~Corresponding real part of the electronic susceptibility. \textbf{d}~Imaginary (left) and real (right) part of the susceptibility using Cs ions for the calculations at $q_z = 0.5$. The red arrow connects the strongest peak in the imaginary part to the corresponding position in the real part, where a smeared-out hump is observed. \textbf{e}~Ratio between the spectral weight of carriers localized by strong electron-phonon coupling and the total intraband spectral weight (Drude + localization peak) at 295~K used as a gauge of electron-phonon coupling.}
\label{Fig4}
\end{figure*}

Fig.~\ref{Fig4}\textbf{a} illustrates the Fermi surfaces at $q_z = 0$ with different colors marking the two bands crossing the Fermi level. To further elucidate changes in the Fermi surface topology, we calculate the static Lindhard susceptibility within the constant matrix element approximation~\cite{Kawamura2019, Johannes2008}:
\begin{equation}
\chi'(\mathbf{q}) = -\sum_{nn'\mathbf{k}} \frac{\theta(E_{\mathrm{F}}-E_{n'\mathbf{k+q}}) - \theta(E_{\mathrm{F}}-E_{n\mathbf{k}})}{E_{n'\mathbf{k+q}} - E_{n\mathbf{k}}}.
\end{equation}
\begin{equation}
\lim_{\omega \to 0}\frac{\chi''(\mathbf{q}, \omega)}{\omega} = \sum_{nn'\mathbf{k}} \delta(E_{n\mathbf{k}} - E_{\mathrm{F}})\delta(E_{n'\mathbf{k+q}} - E_{\mathrm{F}}).
\end{equation}
Here, the imaginary part reflects purely geometrical nesting, while the real part is expected to show a pronounced enhancement at the CDW ordering vector in the case of a Fermi-surface-driven electronic CDW instability. As seen in Fig.~\ref{Fig4}\textbf{b}, strong geometrical nesting is present in the case of Na with $\mathbf{q}_{\mathrm{nest}} = (1/2,1/2,0)$, while in the case of K, multiple weaker and less pronounced nesting vectors are observed. Geometrical nesting is further reduced for Rb, where sharp peaks are completely absent except for the remaining self-nesting at $q = (0,0,0)$. Despite the strong geometrical nesting in the case of Na, only a weak, smeared-out enhancement around $\mathbf{q}_{\mathrm{nest}}$ is observed in  $\chi'(\mathbf{q})$. In the case of Cs and Fr, $\chi''(\mathbf{q})$ develops relatively sharp nesting features with corresponding, although not clearly isolated enhancements in $\chi'(\mathbf{q})$ (see red arrow in Fig.~\ref{Fig4}\textbf{b, c}) at $\mathbf{q}_{\mathrm{nest}} = (1/4,1/4,0)$, which does not correspond to the experimentally observed CDW wavevector~\cite{Kautzsch2023}. While geometrical nesting is generally enhanced at $q_z = 0.5$ as shown in Fig.~\ref{Fig4}\textbf{d} and Supplementary Fig.~S16, changes in $\chi'(\mathbf{q})$ are minor and no well-isolated peaks appear at any wavevector.

From the absence of sharp peaks in the real part of the static Lindhard susceptibilities, we conclude that there is no significant electronic instability associated with Fermi-surface nesting, consistent with previous examinations of the Fermi surface topology of the pristine parent compounds~\cite{Wu2022, Kaboudvand2022} as well as with results in the related ScV$_6$Sn$_6$~\cite{Cao2023, Tan2023}. Instead, an important role of electron-phonon coupling (EPC) has been demonstrated by inelastic x-ray scattering, optical spectroscopy, and first-principles calculations~\cite{Wenzel2023, McGuinness2026, Amigo2024, He2024, Liu2022}. In infrared spectroscopy, strong electron-phonon coupling manifests itself in broad, asymmetric phonon line shapes (see also discussion in Supplementary Note~IIC), as well as in a displaced Drude peak, referred to as the “localization peak” in Fig.~\ref{Fig2}\textbf{a}, which appears in addition to the conventional metallic Drude response. This feature was shown to arise from strong coupling between the electronic and lattice degrees of freedom~\cite{Faria2025, Wenzel2023, Wenzel2025, Wenzel2022}. In \CVS, we explicitly showed that the spectral weight, SW~$= \int \sigma_1(\omega) \mathrm{d}\omega$, of the localization peak provides a measure of electron-phonon coupling strength, which is suppressed by hydrostatic pressure as the CDW vanishes, in agreement with theoretical studies~\cite{Amigo2024, Wenzel2023, Si2022}.

In Fig.~\ref{Fig4}\textbf{e}, we plot the ratio of the spectral weight of the localized carriers to the total intraband spectral weight, comprising the localization and Drude peaks. While the presence of this peak already indicates strong electron-phonon coupling in \AVS\ compounds, we observe pronounced differences among the investigated compounds. The strongest localization contribution is observed for pristine \KVS, while the weakest is observed for \RVS; \CVS\ presents an intermediate case. Overall, the localization contribution evolves continuously with $A$-site composition, with K enhancing and Rb reducing carrier localization, i.e., electron-phonon coupling.

\section{DISCUSSION}
Interestingly, the evolution of the EPC-related localization contribution follows a trend opposite to that of \Tc\ (Fig.~\ref{Fig1}\textbf{e}). \RVS, with the highest \Tc\ of 102~K, shows the weakest carrier localization effects, while the EPC-related localization contribution is strongest for \KVS, which has the lowest \Tc\ of 78~K. This demonstrates that the strength of electron-phonon coupling alone does not determine the CDW transition temperature, suggesting that the CDW transition is governed by a more complex interplay of electronic and lattice degrees of freedom. Soft phonon modes in \AVS\ are shown to be strongly momentum-dependent~\cite{McGuinness2026, Wang2026}. Specifically, ab initio studies considering anharmonic effects reveal enhanced EPC along $M$--$L$ in \CVS~\cite{McGuinness2026}. In Supplementary Fig.~S14, we show that the effect of the band-saddle-point inversion extends along $M$--$L$, suggesting a possible connection between the momentum dependence of the EPC and the orbital character of the electronic states, and consequently, a possible route for tuning electron-phonon interactions in \AVS.

Considering the absence of a Fermi-surface-nesting-driven electronic instability in \AVS\ compounds, the saddle-point inversion together with the strong momentum dependence of the EPC provides a plausible framework for understanding the differences in CDW ordering vector and the nature of CDW. Although alkali-metal states are absent near the Fermi level, the choice of alkali ion systematically controls the low-energy electronic structure across the series. Moreover, the observed evolution cannot be attributed simply to changes in lattice parameters, pointing instead to a modification of the local environment of the kagome layer induced by the alkali ion. One possible microscopic origin is the modification of the local electrostatic potential experienced by the kagome layer. In particular, the large ionic size and electropositive character of Cs may alter the electrostatic environment, particularly the V $3d_{z^2}$ orbital energy. Since the $d_{z^2}$ orbital points out of the plane, its energy should be especially sensitive to these changes, potentially explaining the observed saddle-point inversion and associated modifications in the low-energy electronic structure.

The consequences of this saddle-point tuning may extend beyond the CDW instability to other low-energy properties, including superconductivity. In particular, the evolution of the saddle-point character provides a possible route for modifying the electronic states relevant for superconducting pairing. Whether this contributes to the enhanced superconducting transition temperature and distinct gap structure observed in \CVS\ remains an open question~\cite{}.

More broadly, our results demonstrate that alkali-site substitution provides a clean tuning parameter for modifying the low-energy electronic properties of \AVS\ compounds without directly interfering with the kagome layer or its nearest-neighbor environment. In contrast to V- or Sb-site substitution, this approach preserves the quasi-two-dimensional kagome framework while providing access to the orbital character of the Van Hove singularities, thereby enabling tuning between $m$-type and $p$-type saddle points. The concomitant changes in low-energy interband transitions, Fermi-surface topology, and electron-phonon coupling demonstrate that $A$-site composition provides a direct means of controlling the low-energy electronic properties of \AVS\ compounds.

\section{METHODS}
\subsection{Experimental}
Single crystals of \AVS\ with mixed $A$-site content have been synthesized as explained in Ref.~\cite{Ortiz2023} and their chemical compositions have been determined using energy-dispersive spectroscopy. Prior to the optical measurements, four-contact dc resistivity measurements were performed on the same sample pieces. Temperature- and frequency-dependent reflectivity measurements in the $ab$-plane were carried out on freshly exfoliated samples with the lateral dimensions of approximately 2 $\times$ 3~mm$^2$ and thicknesses on the order of 100~$\mu$m. A broad energy range (50 to 18000~\cm, corresponding to 6 meV -- 2.23~eV) was covered using two Bruker Fourier-transform infrared spectrometers. For low frequencies ($\omega$ < 600~\cm), a IFS113v
spectrometer and a custom-built cryostat were used, while data at higher energies were collected with a Vertex 80v spectrometer coupled to a Hyperion IR microscope. Freshly evaporated gold mirrors served as reference in these measurements and the absolute value of the reflectivity was obtained by the in-situ gold-overcoating technique in the far-infrared range. The complex optical conductivity was calculated via standard Kramers-Kronig analysis~\cite{Dressel2002}, utilizing Hagen-Rubens extrapolations below 50~\cm, and x-ray scattering functions to extrapolate the data to high energies~\cite{Tanner2015}. 
\subsection{Computational}
Density-functional-theory (DFT) calculations of the band structure and optical conductivity were performed in the \texttt{Wien2K} code~\cite{wien2k, Blaha2020} using the Perdew-Burke-Ernzerhof flavor of the exchange-correlation potential~\cite{pbe96}. The lattice parameters of the compounds were determined by x-ray diffraction on polycrystalline alloys. The non-special $z$-coordinate of the Sb2 atoms was interpolated between those of the pure end members (See Supplemental Note~IA). All calculations presented in the main text were performed using the lattice parameters of \KRC\ with $a = b = 5.49782$~\AA, $c = 9.16001$~\AA, and $z_{\mathrm{Sb2}} = 0.74774$. Self-consistent calculations were converged on the 15~$\times$~15~$\times$~8 $k$-mesh. Spin-orbit coupling was included in all calculations if not states otherwise. The Fermi surface was calculated on a $k$-mesh with 26~$\times$~26~$\times$~13, while the optical conductivity was calculated on a denser $k$-mesh with 58~$\times$~58~$\times$~29 points using the OPTIC module~\cite{Draxl2006}.

\section{ACKNOWLEDGEMENTS}
Authors acknowledge the fruitful discussion with Simone Fratini and Erik van Heumen. We are grateful to Gabriele Untereiner for preparing the single crystals for the optical measurements and to Berina Klis for assisting with dc resistivity measurements. M.W. is supported by IQST Stuttgart/Ulm via a project
funded by Carl Zeiss foundation. Work by B.R.O. was supported by the U.S. Department of Energy, Office of Science, Basic Energy Sciences, Materials Sciences and Engineering Division. S.D.W. gratefully acknowledges support via the UC Santa Barbara NSF Quantum Foundry funded via the Q-AMASE-i program under Award No. DMR-1906325. The work has been supported by the Deutsche Forschungsgemeinschaft (DFG) via DR228/51-3, DR228/68-1, and UY63/2-1.

\section{AUTHOR CONTRIBUTIONS}
M.W. performed the experiments and DFT calculations with input from E.U., A.A.T, and M.D. The samples were grown by A.N.C.S., B.R.O., and S.D.W. The manuscript was written by M.W. with suggestions from all authors.

\section{DATA AVAILABILITY}
The data that support the findings of this study are available from the corresponding author upon request.

\section{COMPETING INTERESTS}
The authors declare no financial nor other competing interests.

\nocite{Fratini2014}
\nocite{Rammal2024}
\nocite{Fano1961}
\nocite{Damascelli1997}
%

\end{document}


\widetext
\begin{center}
\textbf{\large Supplementary Information for \\''Band-Saddle-Point Engineering in Mixed $A$-Site \AVS\ Kagome Metals''}

\vspace{5mm}
M. Wenzel, A. A. Tsirlin, A. N. Capa Salinas, B. R. Ortiz, S. D. Wilson, E. Uykur, and M. Dressel

\end{center}
\setcounter{equation}{0}
\setcounter{figure}{0}
\setcounter{table}{0}
\setcounter{page}{1}
\makeatletter
\renewcommand{\theequation}{S\arabic{equation}}
\renewcommand{\thefigure}{S\arabic{figure}}
\renewcommand{\thetable}{S\arabic{table}}
\renewcommand{\bibnumfmt}[1]{[S#1]}
\renewcommand{\citenumfont}[1]{S#1}

\section{Characterization}
\subsection{Lattice Parameters}
\label{parasection}
\begin{figure}[h]
	\centering
	\includegraphics[width=0.8\columnwidth]{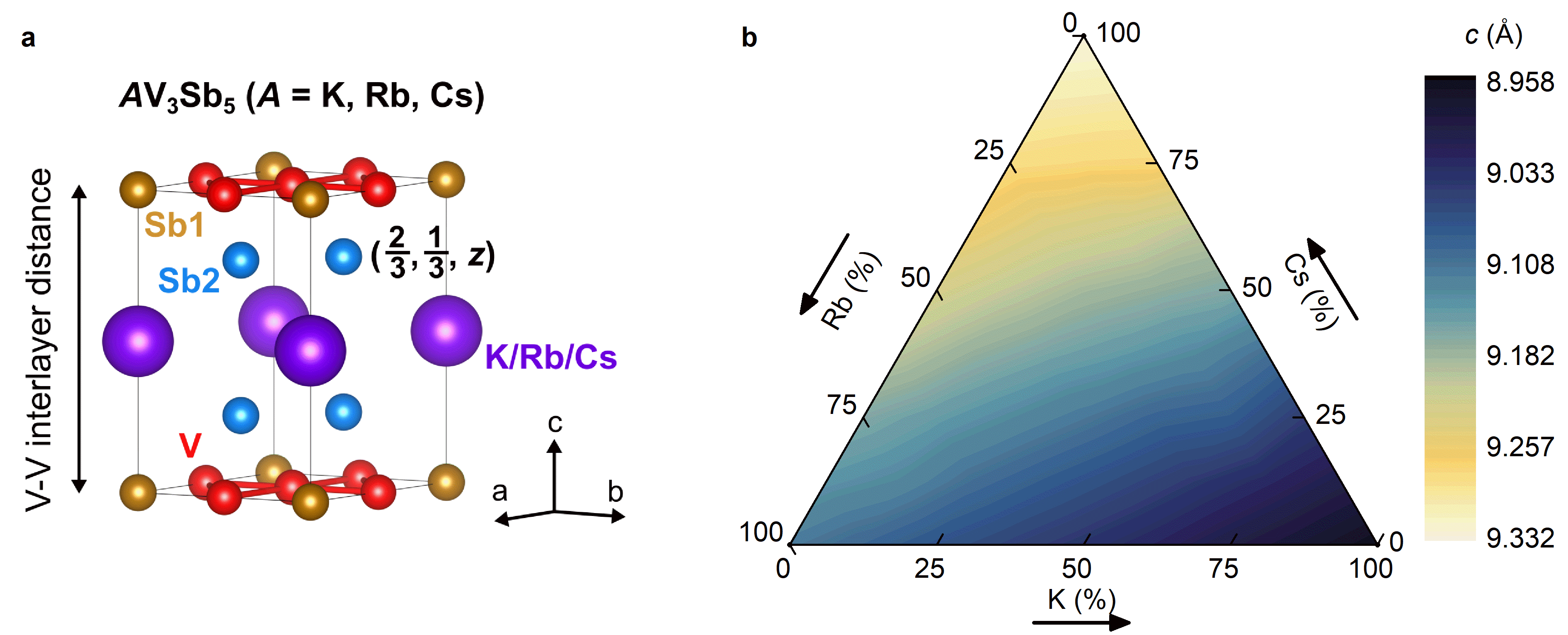}
	\caption{\textbf{a} Crystal structure of \AVS\ highlighting the non-special position of Sb2 atoms, visualized using \texttt{VESTA}~\cite{Momma2008}. \textbf{b}~Evolution of the kagome interlayer distance ($c$ parameter) across the mixed $A$-site compounds. The data were collected on polycrystalline alloys~\cite{Ortiz2023}.}
	\label{latticepara}
\end{figure}

\AVS\ compounds crystallize in the hexagonal $P6/mmm$ space group, with vanadium atoms forming a kagome lattice and Sb1 atoms occupying the centers of the hexagons. These kagome networks are stacked along the $c$-axis and separated by alkali ions and Sb2 honeycomb layers as illustrated in Fig.~\ref{latticepara}\textbf{a}. 

The lattice parameters of mixed $A$-site samples were determined from polycrystalline alloys \cite{Ortiz2023}. Consequently, atomic positions could not be directly refined, and the non-special $z$-coordinates of Sb2 atoms were interpolated based on the pure end members of the \AVS\ compounds, assuming a linear evolution of this parameter. This assumption is reasonable given the nearly linear trend observed for the $c$-axis lattice parameter visualized in Fig.~\ref{latticepara}\textbf{b}. The crystal structure parameters of the mixed $A$-site samples used in the DFT calculations are summarized in Table~\ref{mixedcif}.
\begin{table}[h]
\begin{center}
\begin{tabular}{c|c|c|c|c}
& $a = b$ (\AA)& $c$ (\AA)& $V$ (\AA$^3$)& Sb2 $z$-coordinate \\\hline
K$_{0.25}$Cs$_{0.75}$V$_3$Sb$_5$ &5.50156&9.27737&243.179&0.74501\\
Rb$_{0.52}$Cs$_{0.48}$V$_3$Sb$_5$ &5.50085&9.27740&241.807&0.74616\\
K$_{0.20}$Rb$_{0.43}$Cs$_{0.37}$V$_3$Sb$_5$ &5.49782&9.16001&239.777&0.74774\\
K$_{0.32}$Rb$_{0.68}$V$_3$Sb$_5$ &5.49003&9.05583&236.378& 0.75102\\
K$_{0.81}$Rb$_{0.19}$V$_3$Sb$_5$ &5.48532&8.97850&233.958& 0.75286
\end{tabular}
\caption{Crystal structure parameters of mixed $A$-site compounds used in the DFT calculations.}
\label{mixedcif}
\end{center}
\end{table}

\subsection{DC Resistivity}
\begin{figure}[h]
	\centering
	\includegraphics[width=0.9\columnwidth]{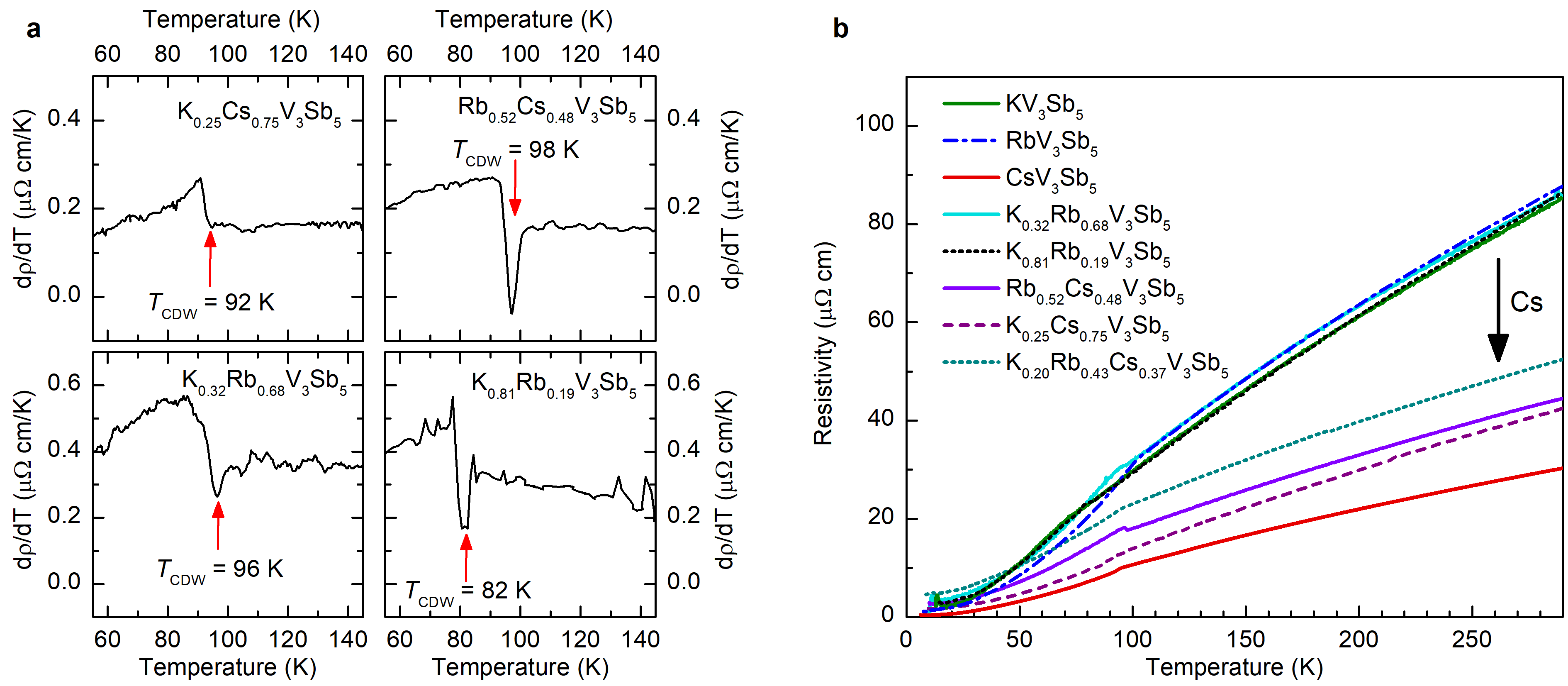}
	\caption{\textbf{a} First derivative of the dc resistivity of mixed $A$-site alloys. The red arrow marks the CDW transition. \textbf{b}~Four-contact dc resistivities measured on the samples used in the optical experiments.}		
	\label{resistivity}
\end{figure}

Prior to the optical studies, four-contact dc resistivity measurements were performed to determine the CDW transition temperature marked by a kink in the dc resistivity. Fig.~\ref{resistivity}\textbf{a} plots the first derivatives which were used to determine \Tc\ presented in the main text. Overall, the resistivity drops with increasing Cs-content as presented in Fig.~\ref{resistivity}\textbf{b}, consistent with the lowest resistivity observed for pristine \CVS. The temperature dependence as well as the absolute values of the dc resistivities match well with the values obtained from the optical measurements via Hagen-Rubens fits of the reflectivities as discussed in Section~\ref{opticssection}.

\begin{table}[h]
\begin{center}
\begin{tabular}{c|c|c|c}
& \Tc\ (K)& $ \Delta_{\mathrm{CDW, absorption}}$ (meV)& $ \Delta_{\mathrm{CDW, peak}}$ (meV)\\\hline
KV$_3$Sb$_5$&78&26.0&77.6\\
RbV$_3$Sb$_5$&102&31.0&85.2\\
CsV$_3$Sb$_5$&94&55.2&101.7\\
K$_{0.25}$Cs$_{0.75}$V$_3$Sb$_5$ &92&52.1&99.5\\
Rb$_{0.52}$Cs$_{0.48}$V$_3$Sb$_5$ &98&44.6&96.4\\
K$_{0.20}$Rb$_{0.43}$Cs$_{0.37}$V$_3$Sb$_5$ &95&38.4&91.3\\
K$_{0.32}$Rb$_{0.68}$V$_3$Sb$_5$ &96&27.9&81.1\\
K$_{0.81}$Rb$_{0.19}$V$_3$Sb$_5$ &82&26.7&79.4
\end{tabular}
\caption{CDW transition temperatures obtained from the kink in dc resistivity, CDW gap energy estimated from the absorption edge ($ \Delta_{\mathrm{CDW, absorption}}$) and from the absorption peak maximum position ($ \Delta_{\mathrm{CDW, peak}}$) as explained in Section~\ref{opticssection}. Values for the parent compounds are determined from the optical data published in Refs.~\cite{Uykur2021, Uykur2022, Wenzel2022}.}
\label{mixedcif}
\end{center}
\end{table}
\section{Optical Spectra}
\label{opticssection}
\begin{figure}[h]
	\centering
	\includegraphics[width=0.9\columnwidth]{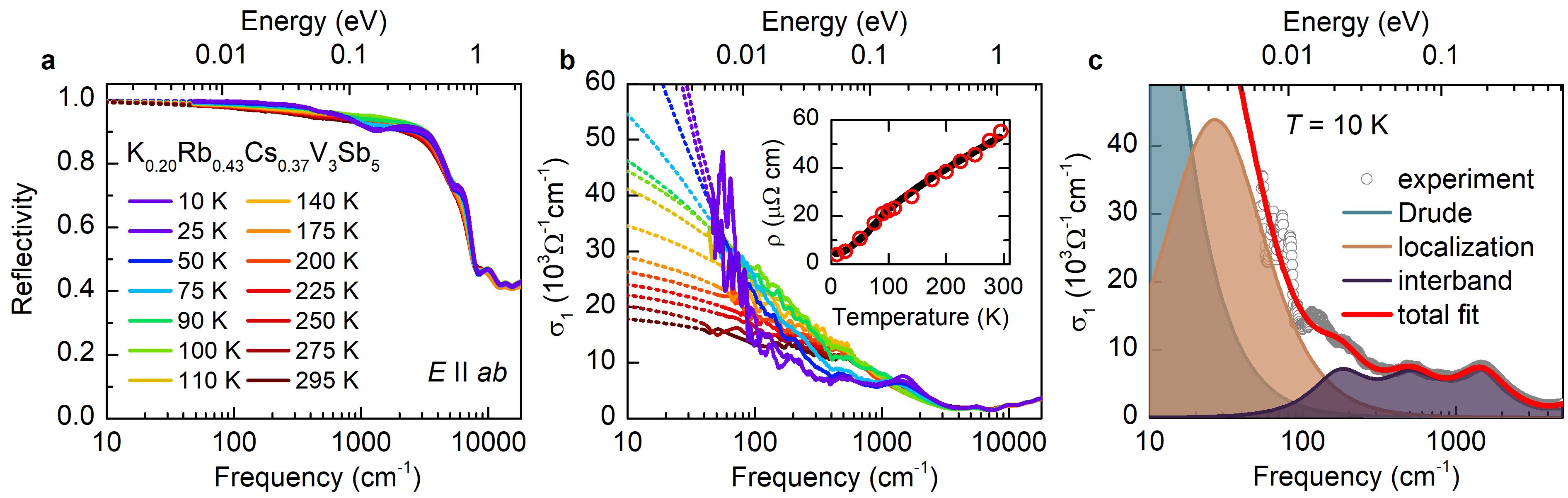}
	\caption{\textbf{a} Temperature-dependent in-plane reflectivity of \KRC\ measured over a broad frequency range. \textbf{b}~Calculated real part of the optical conductivity. The inset displays the four-contact dc resistivity overlapped with the resistivity values obtained from the Hagen-Rubens fits of the reflectivity. \textbf{c} Decomposed optical conductivity of \KRC\ at 10~K modeled by a Drude + localization peak (intraband) and several Lorentzians describing the interband transitions.}		
	\label{KRC_spectra}
\end{figure}

\begin{figure}[!h]
	\centering
	\includegraphics[width=0.95\columnwidth]{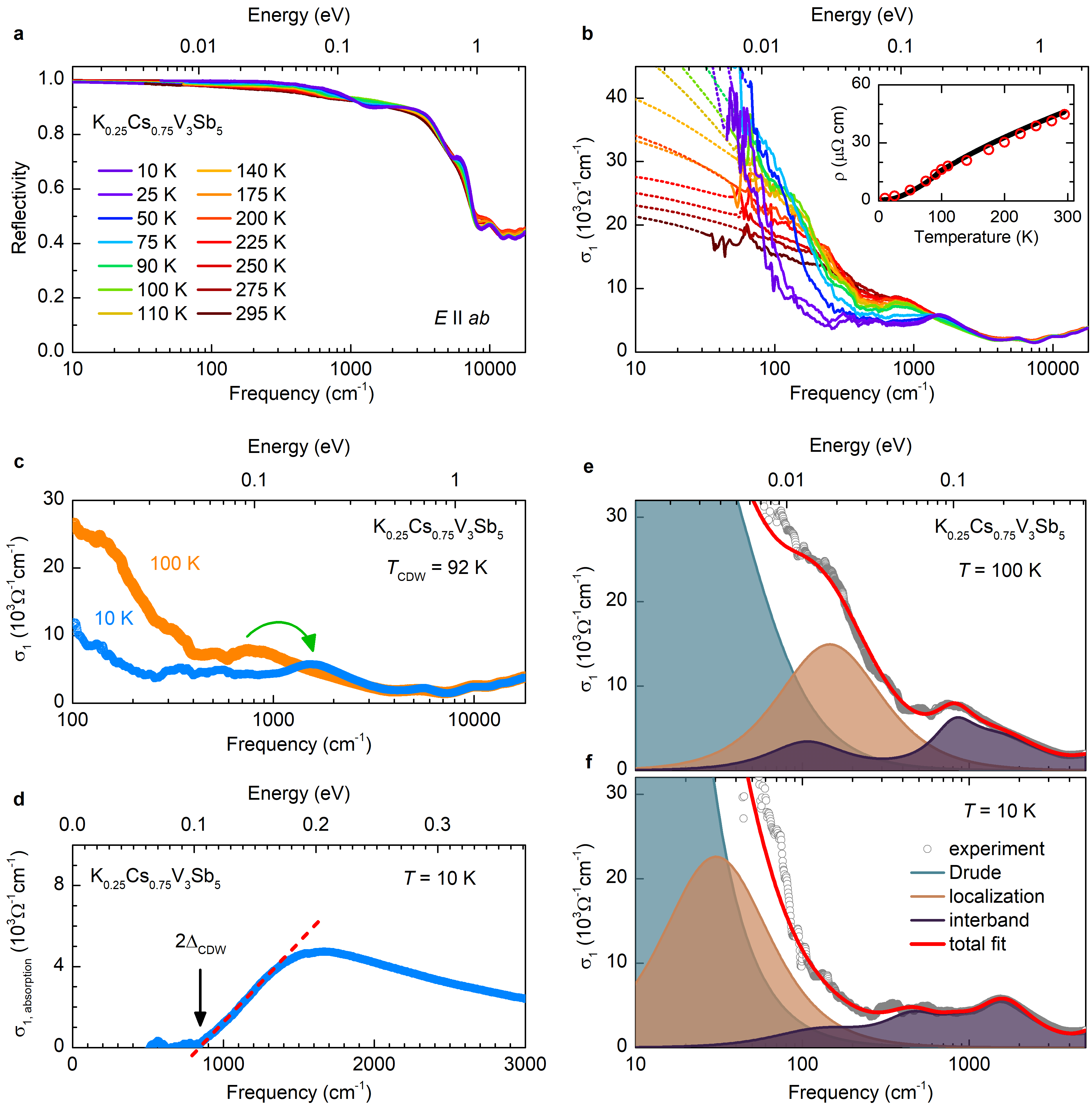}
	\caption{\textbf{a} Temperature-dependent in-plane reflectivity of K$_{0.25}$Cs$_{0.75}$V$_3$Sb$_5$ measured over a broad frequency range. \textbf{b}~Calculated real part of the optical conductivity. The inset displays the four-contact dc resistivity overlapped with the resistivity values obtained from the Hagen-Rubens fits of the reflectivity. \textbf{c} Experimental optical conductivity at 10~K ($T <$ \Tc) and 100~K ($T >$ \Tc). The green arrow
highlights the spectral weight transfer due to the gap opening, leading to a new absorption peak. \textbf{d} Absorption edge after subtracting the low-energy contributions from the spectrum at 10~K. The extrapolated zero crossing (red dashed line) corresponds to 2\DCDW. \textbf{e} Decomposed optical conductivity at 100~K. \textbf{f} Decomposed optical conductivity at 10~K.}		
	\label{KC_spectra}
\end{figure}

\begin{figure}[h]
	\centering
	\includegraphics[width=0.95\columnwidth]{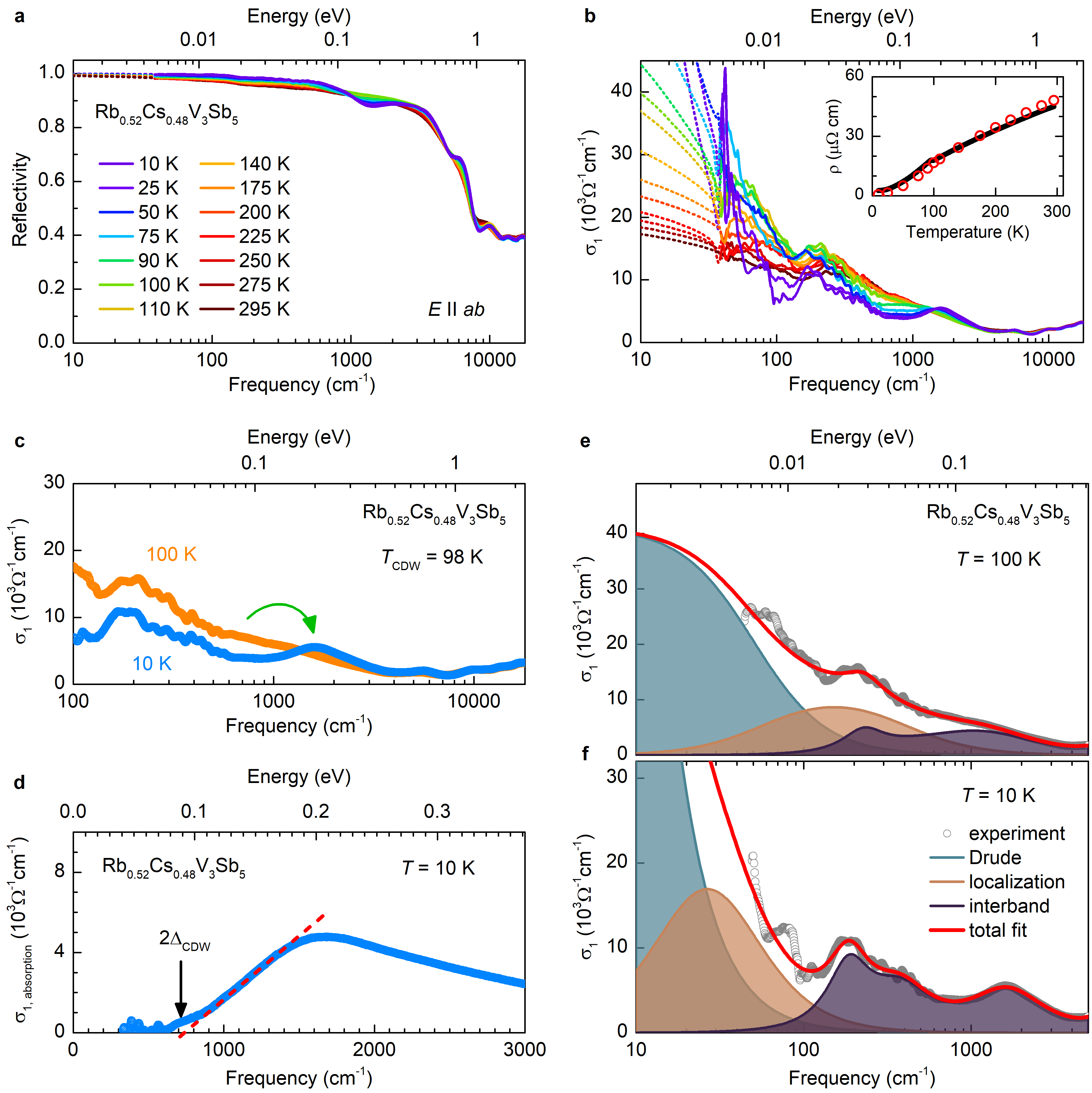}
	\caption{\textbf{a} Temperature-dependent in-plane reflectivity of Rb$_{0.52}$Cs$_{0.48}$V$_3$Sb$_5$ measured over a broad frequency range. \textbf{b}~Calculated real part of the optical conductivity. The inset displays the four-contact dc resistivity overlapped with the resistivity values obtained from the Hagen-Rubens fits of the reflectivity. \textbf{c} Experimental optical conductivity at 10~K ($T <$ \Tc) and 100~K ($T >$ \Tc). The green arrow
highlights the spectral weight transfer due to the gap opening, leading to a new absorption peak. \textbf{d} Absorption edge after subtracting the low-energy contributions from the spectrum at 10~K. The extrapolated zero crossing (red dashed line) corresponds to 2\DCDW. \textbf{e} Decomposed optical conductivity at 100~K. \textbf{f} Decomposed optical conductivity at 10~K.}		
	\label{RC_spectra}
\end{figure}

\begin{figure}[h]
	\centering
	\includegraphics[width=0.95\columnwidth]{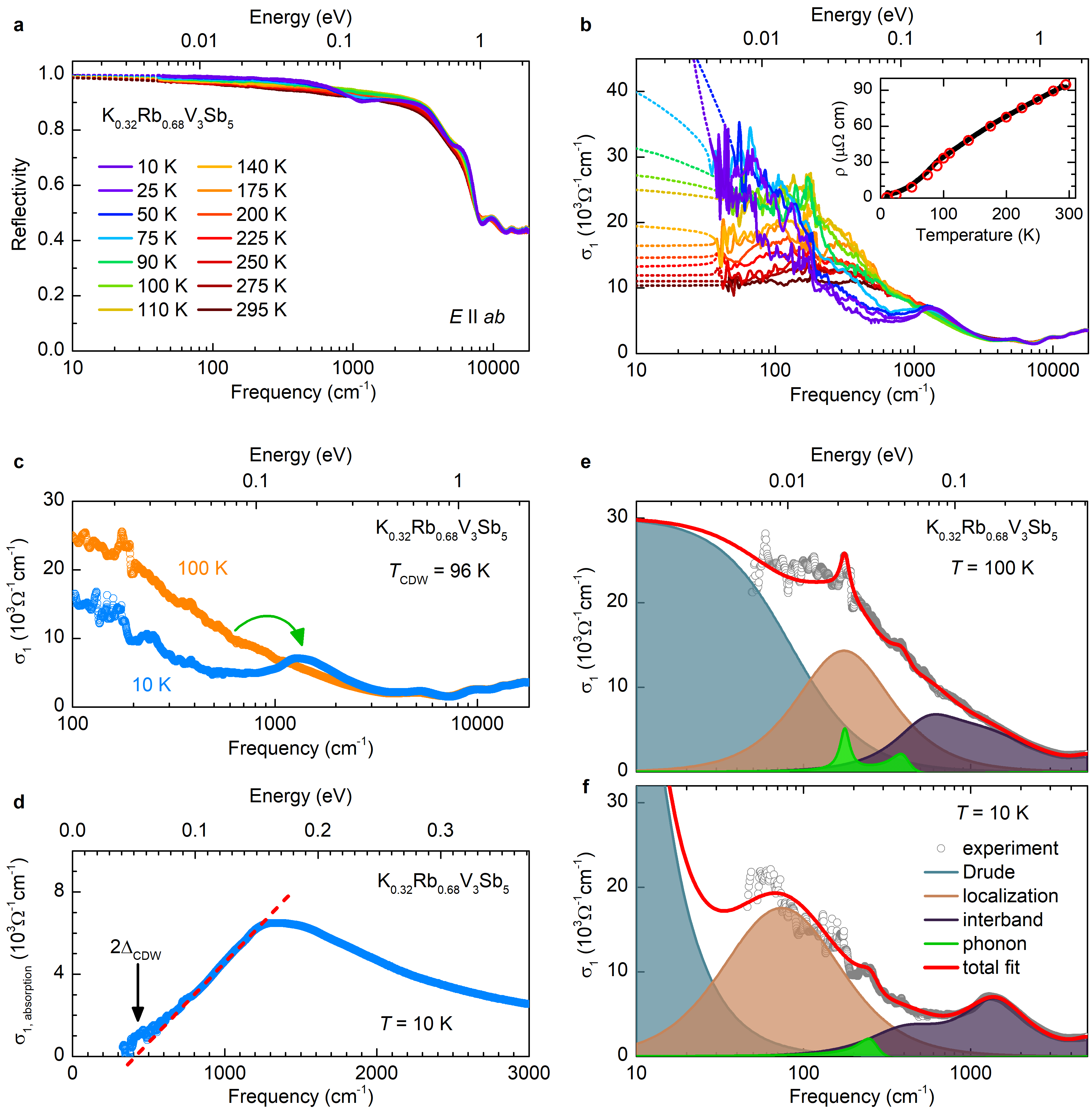}
	\caption{\textbf{a} Temperature-dependent in-plane reflectivity of K$_{0.32}$Rb$_{0.68}$V$_3$Sb$_5$ measured over a broad frequency range. \textbf{b}~Calculated real part of the optical conductivity. The inset displays the four-contact dc resistivity overlapped with the resistivity values obtained from the Hagen-Rubens fits of the reflectivity. \textbf{c} Experimental optical conductivity at 10~K ($T <$ \Tc) and 100~K ($T >$ \Tc). The green arrow
highlights the spectral weight transfer due to the gap opening, leading to a new absorption peak. \textbf{d} Absorption edge after subtracting the low-energy contributions from the spectrum at 10~K. The extrapolated zero crossing (red dashed line) corresponds to 2\DCDW. \textbf{e} Decomposed optical conductivity at 100~K. \textbf{f} Decomposed optical conductivity at 10~K.}		
	\label{K032Rb_spectra}
\end{figure}

\begin{figure}[h]
	\centering
	\includegraphics[width=0.95\columnwidth]{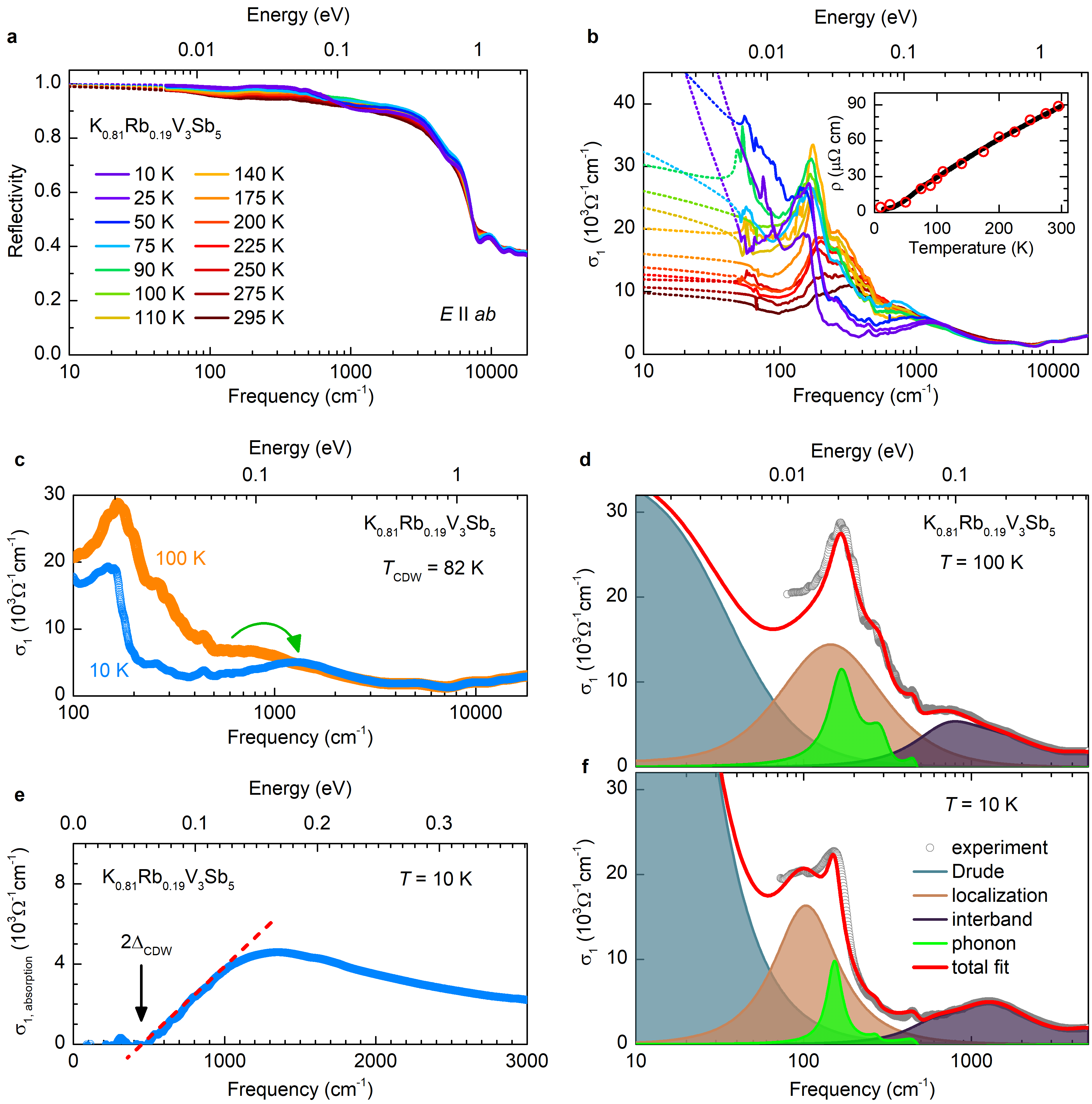}
	\caption{\textbf{a} Temperature-dependent in-plane reflectivity of K$_{0.81}$Rb$_{0.19}$V$_3$Sb$_5$ measured over a broad frequency range. \textbf{b}~Calculated real part of the optical conductivity.The inset displays the four-contact dc resistivity overlapped with the resistivity values obtained from the Hagen-Rubens fits of the reflectivity. \textbf{c} Experimental optical conductivity at 10~K ($T <$ \Tc) and 100~K ($T >$ \Tc). The green arrow
highlights the spectral weight transfer due to the gap opening, leading to a new absorption peak. \textbf{d} Absorption edge after subtracting the low-energy contributions from the spectrum at 10~K. The extrapolated zero crossing (red dashed line) corresponds to 2\DCDW. \textbf{e} Decomposed optical conductivity at 100~K. \textbf{f} Decomposed optical conductivity at 10~K.}		
	\label{K081Rb_spectra}
\end{figure}

This section presents the raw reflectivity and optical conductivity data of the mixed $A$-site samples. The temperature-dependent reflectivities measured in the $ab$-plane are shown in panels \textbf{a} of Figs. \ref{KRC_spectra}--\ref{K081Rb_spectra}, with the corresponding optical conductivities given in panels \textbf{b}. The insets show the four-contact dc resistivity measurements overlapped with the values obtained from the Hagen-Rubens extrapolations.

\begin{figure}[h]
	\centering
	\includegraphics[width=0.9\columnwidth]{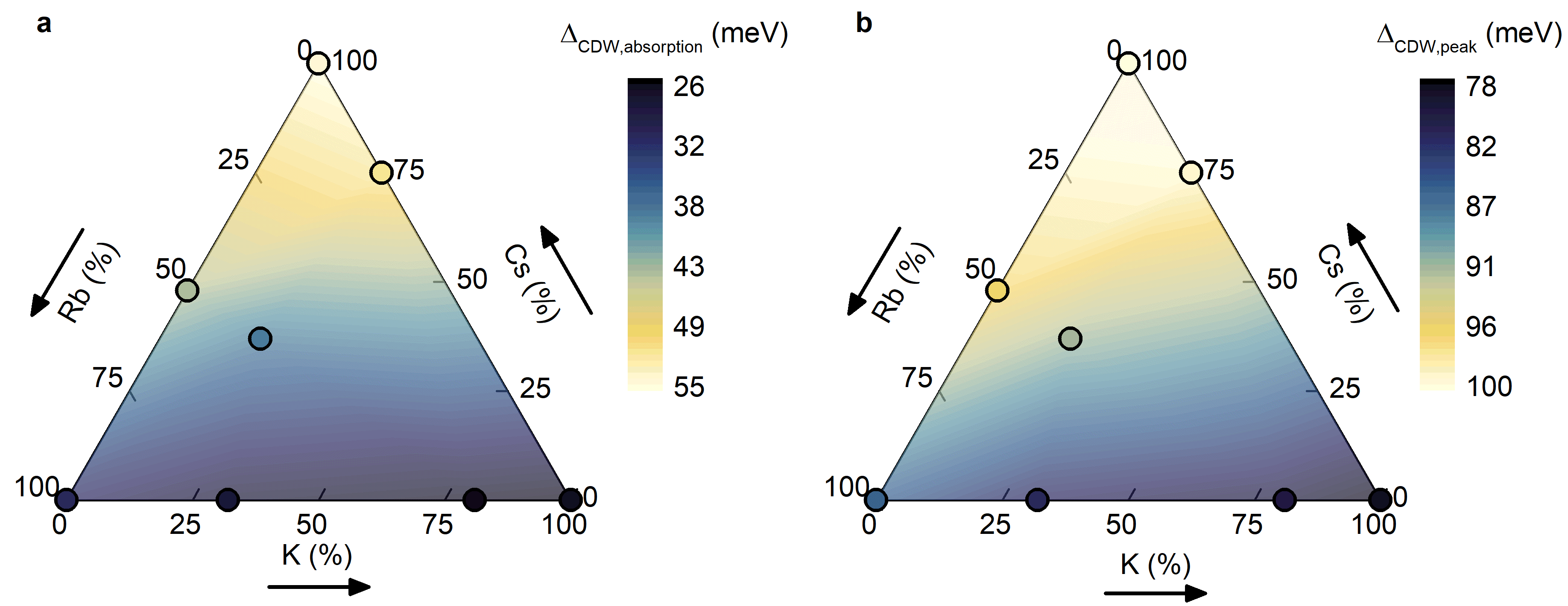}
	\caption{Alkali-site-content-evolution of the CDW gap energy determined from the absorption edge (\textbf{a}) and from the position of the absorption peak
maximum (\textbf{b}). Experimental values are taken at 10~K and are highlighted with colored dots.}		
	\label{CDWgap}
\end{figure}

Upon cooling below \Tc, significant changes occur in the optical spectra of all mixed $A$-site samples. A dip around 0.15~eV begins to form in the reflectivity, which is echoed by a spectral-weight transfer observed in the optical conductivity as indicated by the green arrow in Figs.~\ref{KC_spectra}--\ref{K081Rb_spectra}\textbf{c} and in Fig.~1\textbf{c} in the main text. The CDW gap energy is estimated either directly from the position of the newly formed absorption peak (green arrow) or by decomposing the spectra, and subtracting all low-energy contributions unrelated to the CDW absorption peak resulting in a prominent absorption edge shown in panels \textbf{d}. This edge is then extrapolated to $\sigma_1 = 0$, with the zero crossing marking the value of 2\DCDW. Both method reveal a continuous but non-linear trend in \DCDW, as demonstrated in Fig.~\ref{CDWgap}.

\subsection{Decomposition}
Different contributions to the total optical spectra are modeled with the Drude-Lorentz approach. The dielectric function [$\tilde{\varepsilon}=\varepsilon_1 + i\varepsilon_2$] is expressed as
\begin{equation}
\label{Eps}
\tilde{\varepsilon}(\omega)= \varepsilon_\infty - \frac{\omega^2_{p,{\rm Drude}}}{\omega^2 + i\omega/\tau_{\rm\, Drude}} + \sum\limits_j\frac{\Omega_j^2}{\omega_{0,j}^2 - \omega^2-i\omega\gamma_j},
\end{equation}
with $\varepsilon_{\infty}$ being the high-energy contributions to the real part of the dielectric permittivity. The Drude parameters $\omega_{p,{\rm Drude}}$ and $1/\tau_{\rm\,Drude}$ describe the plasma frequency and the scattering rate of the itinerant carriers, respectively. Lorentzians with the resonance frequency $\omega_{0,j}$, the strength of the oscillation $\Omega_j$, and the width $\gamma_j$ are used to model the interband absorptions and symmetric phonon modes.

The complex optical conductivity [$\tilde{\sigma}=\sigma_1 + i\sigma_2$] is calculated as
\begin{equation}
\tilde{\sigma}(\omega)= -i\omega \varepsilon_{\mathrm{0}}[\tilde{\varepsilon} (\omega) -\varepsilon_\infty],
\end{equation}
with $\varepsilon_{\mathrm{0}}$ being the vacuum permittivity.

In addition to the conventional metallic Drude response, the optical spectra of kagome metals are characterized by the presence of a displaced Drude peak, independent of the band filling and magnetic structure~\cite{Uykur2021, Uykur2022, Wenzel2022, Wenzel2025, Faria2025}. This feature represents carriers with hindered dynamics and can be described theoretically by the presence of strong electron-phonon interactions leading to~\cite{Fratini2014, Rammal2024}.
\begin{equation}
\label{Fratini}
\tilde{\sigma}_{\rm localization}(\omega)=\frac{C}{\tau_{\mathrm{b}}-\tau}\frac{\tanh\{\frac{\hbar\omega}{2k_{\mathrm{B}}T}\}}{\hbar\omega} \cdot\,\left\{\frac{1}{1-\mathrm{i}\omega\tau}-\frac{1}{1-\mathrm{i}\omega\tau_{\mathrm{b}}}\right\},
\end{equation}
where $C$ is a constant, $\hbar$ is the reduced Planck constant, $k_{\mathrm{B}}$ the Boltzmann constant, $\tau$ the elastic scattering time of the standard Drude model, and $\tau_{\mathrm{b}}$ the backscattering time leading to localization. 

Asymmetric phonon modes are modeled by the Fano model \cite{Fano1961, Damascelli1997}
\begin{equation}
\tilde{\sigma}(\omega) = \mathrm{i}\sigma_0\frac{(q - \mathrm{i})^2}{\mathrm{i} + \Omega},\,\,\,\,\,\,\Omega = \frac{\omega^2 - \omega_0^2}{\gamma \omega}
 \label{fano}
\end{equation}
Here, $\sigma_{\mathrm{0}}$ is the resonance strength, $\gamma$ the damping and $\omega_{\mathrm{0}}$ the resonance frequency. The shape of the Fano resonance highly depends on the dimensionless parameter $q$, related to the phase shift between the discrete mode (phonon) and the electronic continuum. The typical asymmetric line shape is obtained for moderate values of $|q|$, while for $|q| \rightarrow \infty$, the line evolves into a symmetric Lorentzian. For $q = 0$, a strong symmetric anti-resonance is observed.

The total optical conductivity takes the form
\begin{equation}
\tilde{\sigma}(\omega)= \underbrace{\tilde{\sigma}_{\mathrm{Drude}} + \tilde{\sigma}_{\mathrm{localization}}}_{\mathrm{intraband}} + \underbrace{\tilde{\sigma}_{\mathrm{Fano}}}_{\mathrm{phonons}} + \underbrace{\tilde{\sigma}_{\mathrm{Lorentzians}}}_{\mathrm{phonons + interband}}.
\end{equation}
The robustness of the fits is ensured by fitting the reflectivity, and the complex optical conductivity simultaneously.

\subsection{Localized Electrons}
In addition to the differences in low-energy interband transitions discussed in the main text, the choice of alkali ion significantly affects the intraband response. Among the parent systems, the response of localized carriers is most pronounced in \KVS, while the broader Drude response and presence of low-energy interband transitions in the case of \CVS\ partially mask this feature in \RVS\ and \CVS~\cite{Uykur2021, Uykur2022, Wenzel2022}. The decomposed optical conductivities of the mixed $A$-site samples given in Fig.~2\textbf{a} of the main text, Fig.~\ref{KRC_spectra}\textbf{c}, and Figs.~\ref{KC_spectra}--\ref{K081Rb_spectra}\textbf{e,f}, confirm this trend. Localization effects generally become more pronounced with increasing potassium content; concomitantly, the Drude response sharpens, signaling a transfer of spectral weight from free to localized electrons as discussed in the main text.

\begin{figure}[h]
	\centering
	\includegraphics[width=0.9\columnwidth]{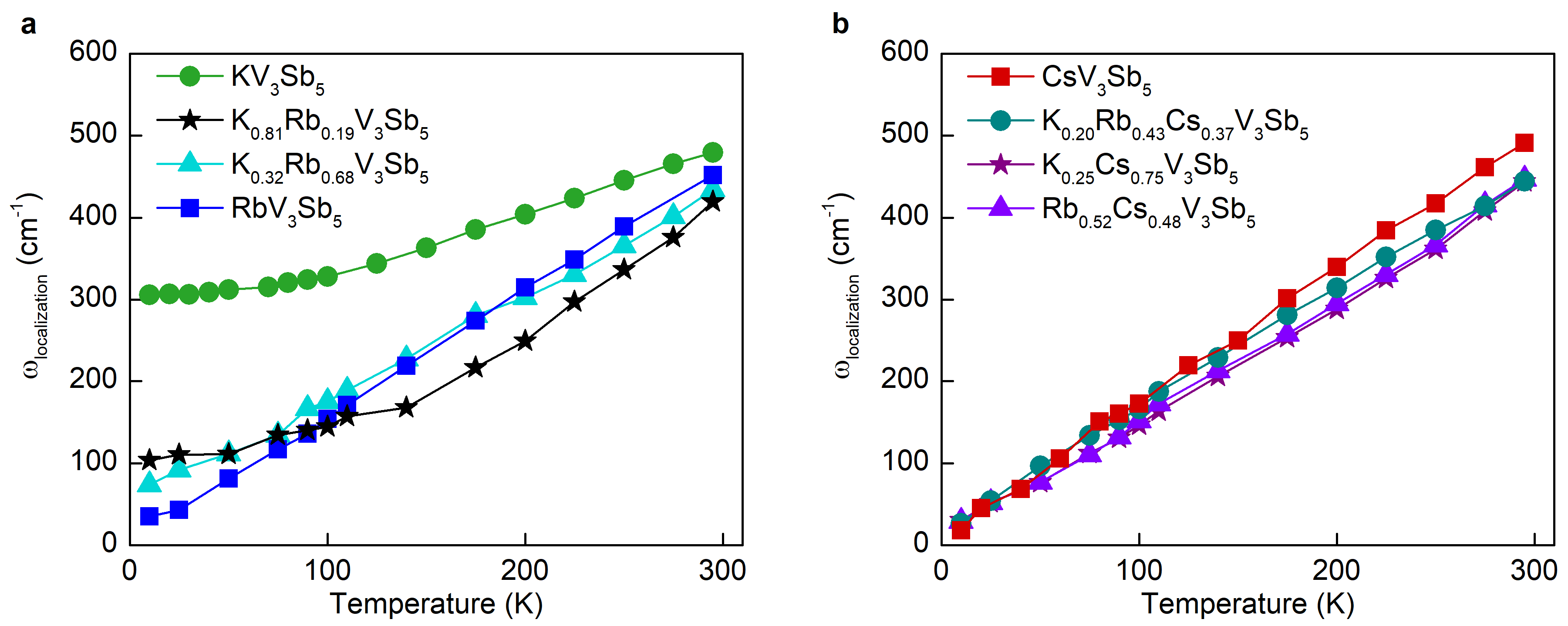}
	\caption{Position of the localization peak as a function of temperature. Panel \textbf{a} focuses on Cs-free samples, where, at low temperatures, a saturation of the peak position is observed for samples with high potassium content. Panel \textbf{b} displays the peak position for Cs-containing samples
showing a linear redshift down to the lowest temperatures. The data of the parent compounds are taken from Refs.~\cite{Uykur2021, Uykur2022, Wenzel2022}.}		
	\label{localization}
\end{figure}

Upon cooling, the localization peak shifts to lower energies and sharpens. The position of the localization peak as a function of temperature for pure and mixed $A$-site compounds is summarized in Fig.~\ref{localization}. At higher temperatures, the peak experiences a linear redshift in all samples. For pristine \RVS\ and all Cs-containing compounds, this trend persists down to the lowest measured temperatures, where eventually, the localization peak merges with the Drude response (see decomposed optical conductivities at 10~K). In contrast, a saturation of the peak position is observed in \KVS\ at temperatures below \Tc~\cite{Uykur2022}. A similar saturation is also observed for K$_{0.81}$Rb$_{0.19}$V$_3$Sb$_5$ as shown in Fig.~\ref{localization}\textbf{a}. Rb-doping gradually transforms this behavior into a linear shift down to the lowest temperatures.

\subsection{Phonon Modes}
The optical spectra of the Cs-free compounds reveal multiple low-energy phonon modes which closely resemble those observed in pristine \KVS\ and \RVS~\cite{Uykur2022, Wenzel2022}. The spectra of pristine \KVS\ and \RVS\ show an IR-active $E_{1u}$ mode at 190~\cm\ and 160~\cm, respectively, as well as a second, higher-energy mode at 480~\cm\ and 430~\cm. The latter modes cannot be readily assigned to IR-active $\Gamma$-point phonons, since no such modes are expected above 250~\cm. They exhibit a pronounced asymmetric line shape that is best described by a Fano resonance rather than a simple Lorentz oscillator.

\begin{figure}[!h]
	\centering
	\includegraphics[width=0.85\columnwidth]{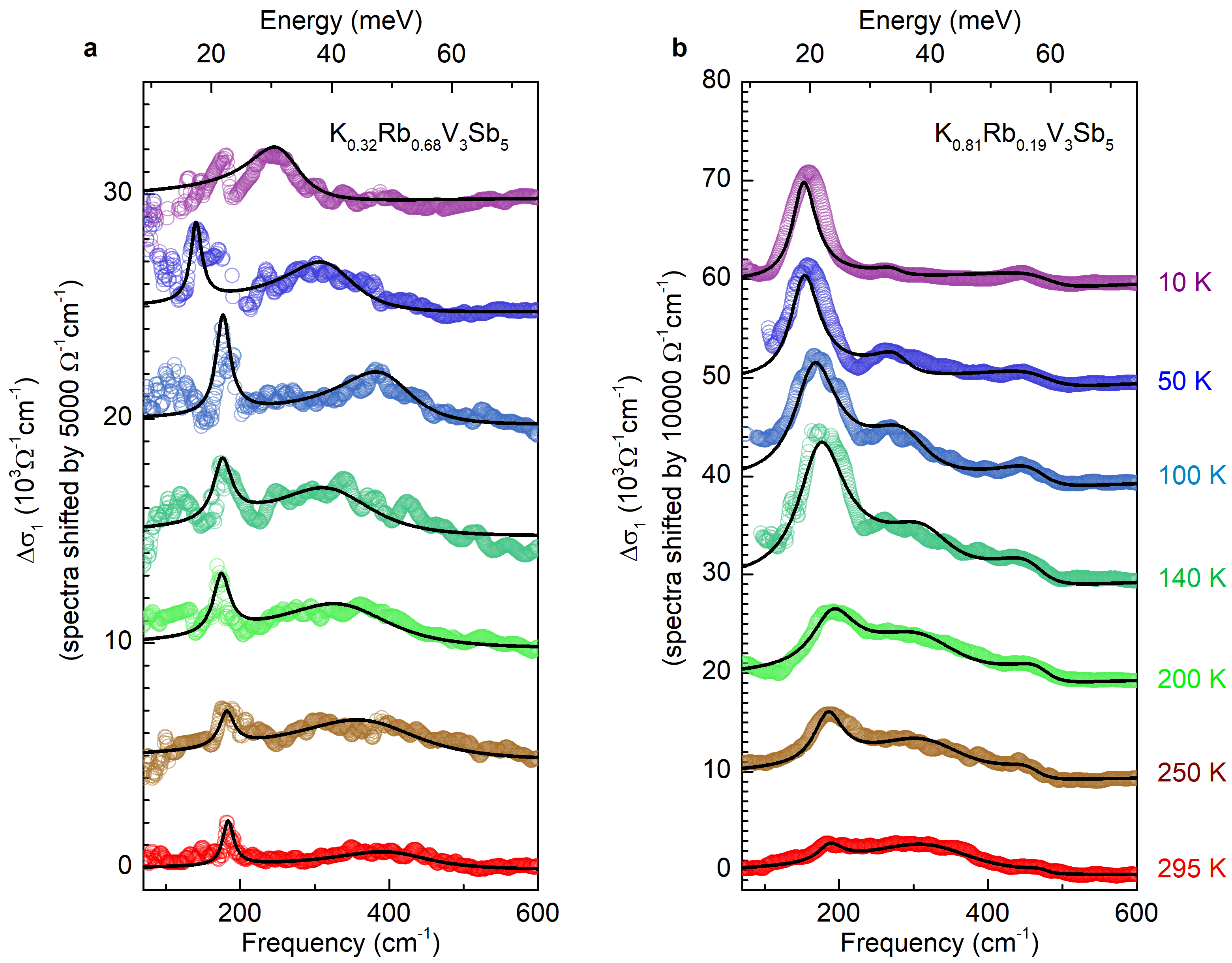}
	\caption{Low-energy optical conductivity of K$_ {0.32}$Rb$_{0.68}$V$_3$Sb$_5$ (\textbf{a}) and K$_ {0.81}$Rb$_{0.19}$V$_3$Sb$_5$ (\textbf{b}) at selected temperatures. The spectra are presented as $\Delta \sigma_1(\omega)$ with all other contributions subtracted and the baselines shifted by 5000 $\Omega^{-1}$\cm\ (\textbf{a}) and 10000 $\Omega^{-1}$\cm\ (\textbf{b}) for clarity. Black lines represent the fits to the optical spectra.}		
	\label{phononfit}
\end{figure}

\begin{figure}[!h]
	\centering
	\includegraphics[width=0.8\columnwidth]{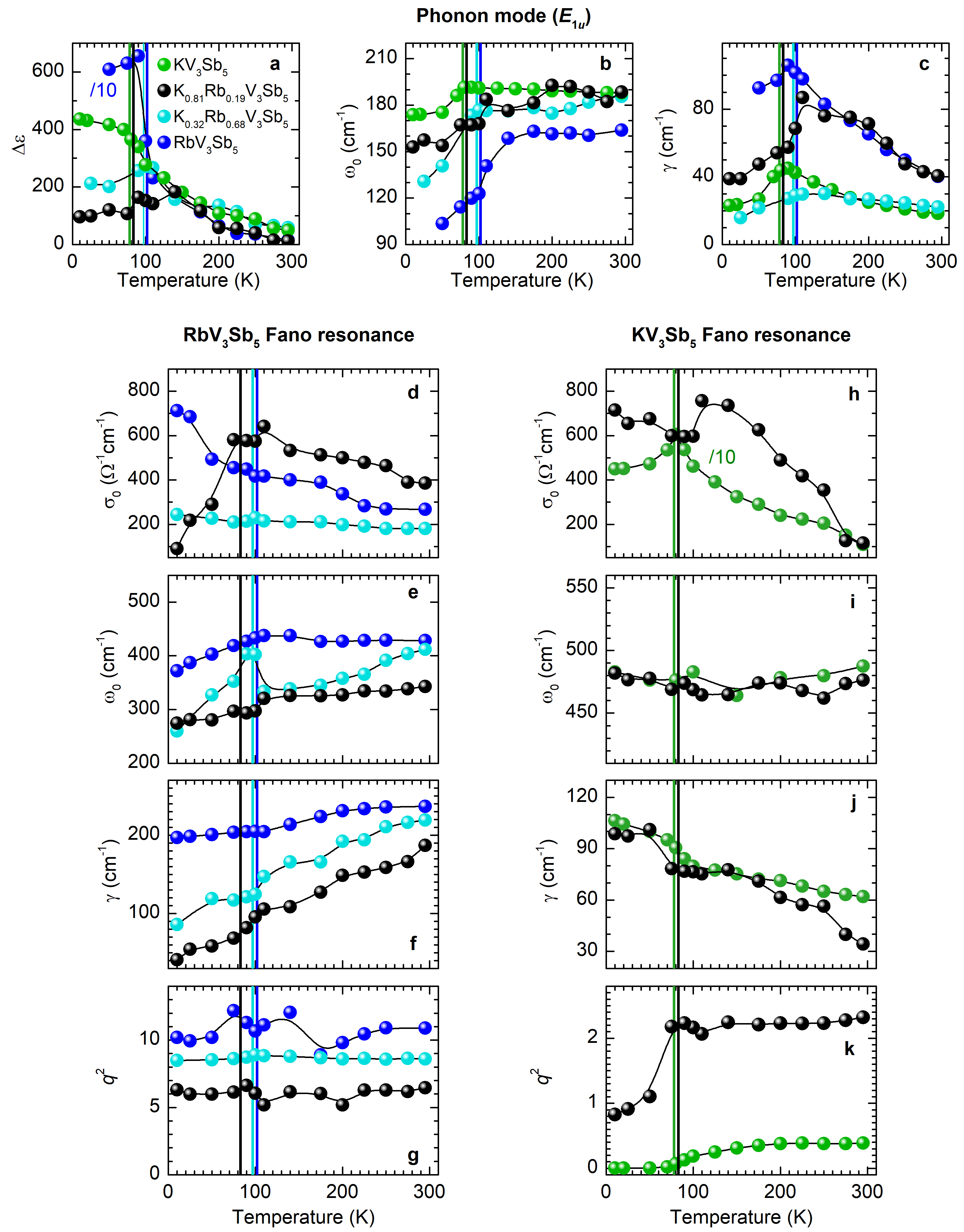}
	\caption{\textbf{a-c} Lorentzian fit parameters of the $E_{1u}$ mode in Cs-free mixed $A$-site compounds. \textbf{d-g} Fit parameters of the Fano resonance, showing similarities to the Fano mode in pristine \RVS. \textbf{h-k} Fit parameters of the higher-energy Fano resonance observed in K$_{0.81}$Rb$_{0.19}$V$_3$Sb$_5$ (black), compared to the Fano mode in pristine \KVS\ (green). Vertical solid lines indicate the CDW transition temperatures. The parameters of the pristine compounds are taken from Refs.~\cite{Uykur2022, Wenzel2022}.}		
	\label{phononpara}
\end{figure}

Similar phonon features are observed in K$_{0.32}$Rb$_{0.68}$V$_3$Sb$_5$ and  K$_{0.81}$Rb$_{0.19}$V$_3$Sb$_5$, as shown in Fig.~\ref{phononfit}. The conventional low-energy $E_{1u}$ phonon is modeled by a Lorentz oscillator. The fit parameters, presented in Fig.~\ref{phononpara}\textbf{a-c}, reveal a pronounced broadening upon cooling, indicative of strong electron-phonon coupling. In addition, the oscillator strength $\Delta\varepsilon$ increases as the CDW transition is approached, followed by a pronounced redshift below \Tc. This anomalous softening upon cooling contrasts with the conventional behavior of phonon modes, which typically harden due to lattice stiffening.

At room temperature, the spectra exhibit a relatively broad Fano resonance at approximately 410~\cm\ for K$_{0.32}$Rb$_{0.68}$V$_3$Sb$_5$ and 350~\cm\ for K$_{0.81}$Rb$_{0.19}$V$_3$Sb$_5$. The temperature dependence of the fit parameters, given in Fig.~\ref{phononpara}\textbf{d-g}, closely resembles that of the Fano mode in pristine \RVS. In particular, the resonance sharpens upon cooling, while the Fano parameter $q$ remains nearly constant.

For K$_{0.81}$Rb$_{0.19}$V$_3$Sb$_5$, an additional Fano resonance is observed at slightly higher energy (Fig.~\ref{phononfit}\textbf{b}). This mode closely follows the behavior of the corresponding Fano resonance in pristine \KVS, with an almost temperature-independent resonance frequency, a pronounced drop in $q^2$ at \Tc, and significant broadening upon cooling, as shown in Fig.~\ref{phononpara}\textbf{h-k}.

The absence of clearly resolved phonon modes in the optical spectra of the Cs-containing compounds does not necessarily indicate a substantial modification of their phonon spectra or the absence of electron-phonon coupling. Instead, it can be attributed to their lower dc resistivity, and consequently, stronger screening of IR-active phonon modes by the conduction electrons.

\clearpage
\newpage

\section{Additional computational results}

\subsection{Orbital-resolved Band Structures}

\begin{figure}[h]
	\centering
	\includegraphics[width=1\columnwidth]{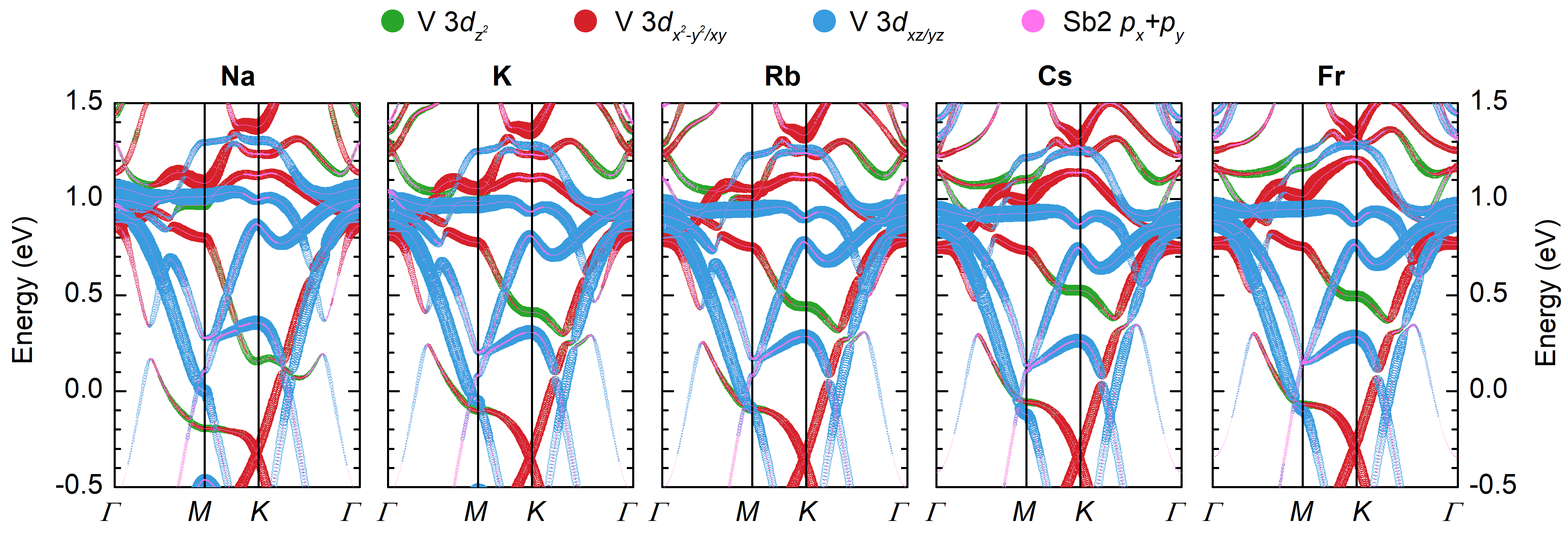}
	\caption{Low-energy band structures of \AVS\ compounds along $\Gamma$--$M$--$K$--$\Gamma$ with colored dots representing contributions from different atomic orbitals. Experimental lattice parameters of \KRC\ were used for all calculations, while the alkali ion was varied from Na to Fr (left to right).}		
	\label{fatBZ1}
\end{figure}

\begin{figure}[h]
	\centering
	\includegraphics[width=1\columnwidth]{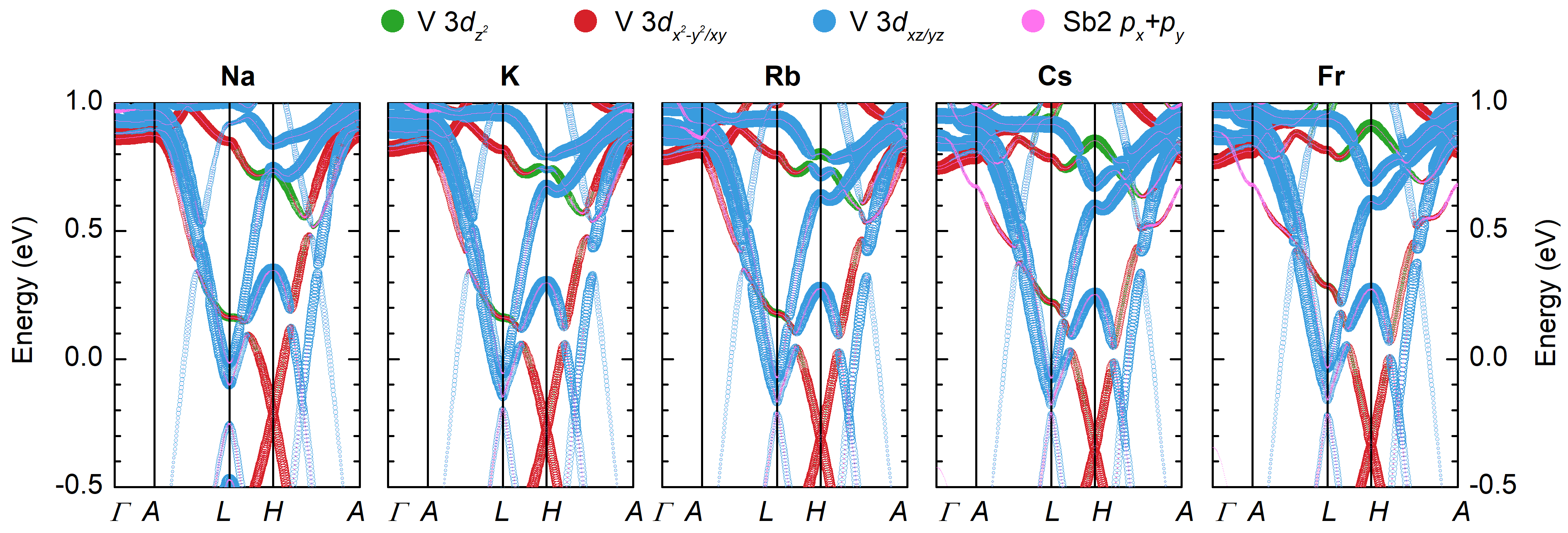}
	\caption{Low-energy band structures of \AVS\ compounds along $\Gamma$--$A$--$L$--$H$--$A$ with colored dots representing contributions from different atomic orbitals. Experimental lattice parameters of \KRC\ were used for all calculations, while the alkali ion was varied from Na to Fr (left to right).}		
	\label{fatBZ2}
\end{figure}

\begin{figure}[h]
	\centering
	\includegraphics[width=1\columnwidth]{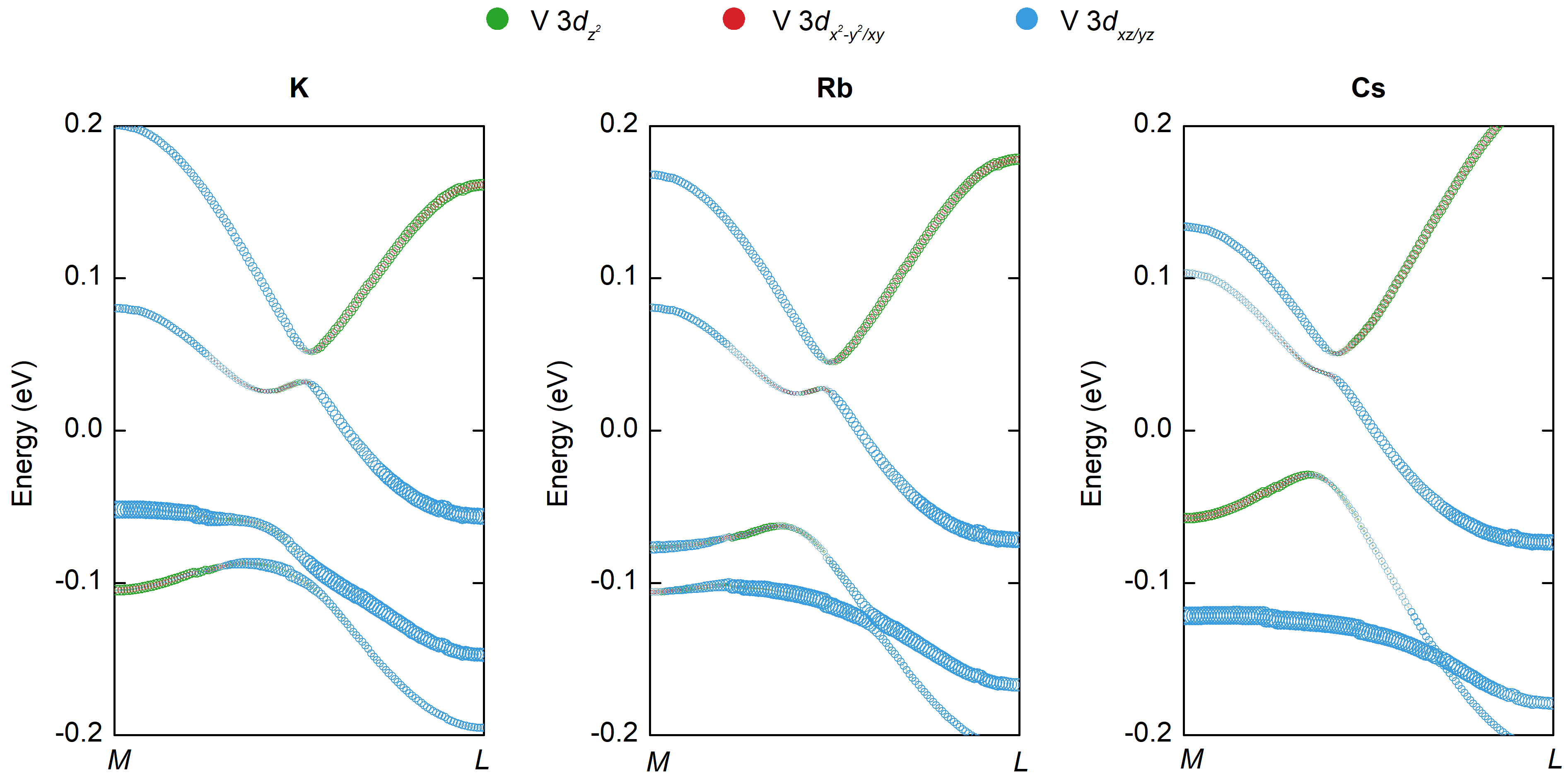}
	\caption{Low-energy band structures of \AVS\ compounds along $M$--$L$ with colored dots representing contributions from different atomic orbitals. Experimental lattice parameters of \KRC\ were used for all calculations, while the alkali ion was varied from K to Cs (left to right).}		
	\label{fatML}
\end{figure}

Figs.~\ref{fatBZ1}--\ref{fatML} show the orbital-resolved band structures along different $k$-directions calculated using the lattice parameters of \KRC. As discussed in the main text, the main modification of the low-energy electronic structure concerns the inversion of the van Hove singularities at $M$, which can be understood as an upward shift of the 'green' vanadium $3d_{z^2}$ band. This shift can also be traced in vicinity of other high-symmetry $k$-points, such as $K$ (see Fig.~\ref{fatBZ1}), $L$, and $H$ (see Fig.~\ref{fatBZ2}). In particular, strongly momentum-dependent electron-phonon coupling has been reported from anharmonic phonon calculations in pristine \CVS\ along the $M$--$L$ direction, associated with the soft mode instability driving the CDW~\cite{McGuinness2026}. In Fig.~\ref{fatML}, we plot the orbital-resolved band structure along this $k$-direction, showing that the effects of the band saddle-point inversion at $M$ extend along $M$--$L$. This suggests that orbital-selective electron-phonon coupling may contribute to the unique CDW properties of \CVS, enabled by the band saddle-point inversion.

\subsection{Interband Transitions}
\begin{figure}[h]
	\centering
	\includegraphics[width=0.9\columnwidth]{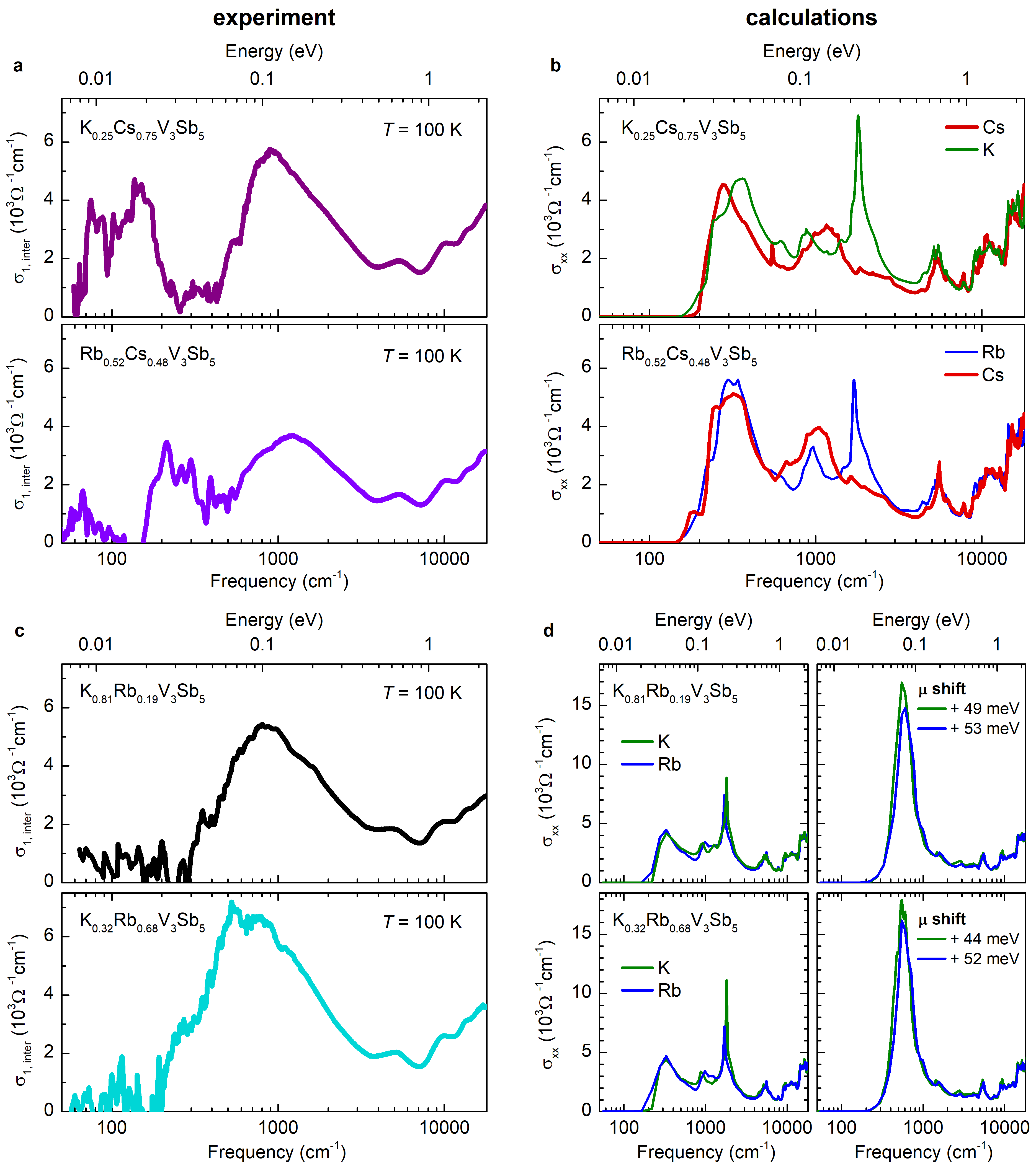}
	\caption{\textbf{a}~Experimental interband optical conductivity of Cs-containing mixed $A$-site compounds obtained by subtracting the Drude and localization peaks from the optical conductivity at 100 K ($T >$~\Tc). \textbf{b}~In-plane component of the optical conductivity derived by DFT in the
normal state. Different alkali ions were used in the calculations as explained
in the text. \textbf{c}~Experimental interband optical conductivity of Cs-free mixed $A$-site compounds at 100 K ($T >$~\Tc). \textbf{d}~In-plane component of the DFT optical conductivity of Cs-free compounds. A better agreement with the experiments is achieved by increasing the total number of electrons by +0.5$e$/f.u., which results in an upward shift of the Fermi level expressed by the chemical potential~$\mu$.}		
	\label{interband}
\end{figure}

Here, we present DFT calculations of the optical conductivity performed using the experimental lattice parameters of the specific mixed $A$-site samples, as described in Section~\ref{parasection}. The $A$-site atom was varied in the calculations while keeping the experimental lattice parameters fixed. The results shown in Fig.~\ref{interband} confirm that variations in the lattice parameters have only a minor impact on the calculated interband optical conductivity. Importantly, in agreement with the calculations presented in the main text, the experimental low-energy interband transitions of the Cs-containing compounds are reproduced only when Cs occupies the $A$-site. Replacing Cs with K or Rb in the calculations results in a sharp absorption peak near 2000~\cm, in clear disagreement with the experimental findings.

For the Cs-containing compounds, the overall agreement between DFT and experiment indicates that the electronic structure is well described by band theory. Nevertheless, small quantitative discrepancies remain in the absorption intensity and at very low energies, where experimental interband transitions appear at slightly lower energies than predicted by the DFT calculations. The latter discrepancy can be accounted for by a modest rescaling of the energy axis, which shifts the calculated interband transitions to lower energies, consistent with a moderate renormalization of the band energies.

In contrast, the optical spectra of pristine \KVS\ and \RVS\ are not well captured by simple band theory. Here, the agreement can be substantially improved by applying a rigid upward shift of the Fermi level, indicating more pronounced orbital-selective band renormalization effects~\cite{Uykur2022, Wenzel2022}. A similar upward shift of the Fermi level improves the agreement between the calculated and experimental interband optical conductivity of the Cs-free mixed $A$-site compounds (see Fig.~\ref{interband}\textbf{c,d}). Here, a moderate shift, corresponding to the addition of half an electron per formula unit, successfully reproduces the main low-energy absorption peak centered around 1000~\cm\ observed experimentally. Since no scattering is included in the calculations, the intensity of the calculated interband absorption is significantly higher than that observed experimentally.

Note that these adjustments are used solely to improve the match between the experimental and DFT-derived interband transitions and do not modify the underlying band dispersion or orbital character of the calculated electronic states. In particular, they do not affect the discussion on the differences associated with the band saddle-point inversion.

\clearpage
\subsection{Fermi Surface}
\begin{figure}[h]
	\centering
	\includegraphics[width=0.65\columnwidth]{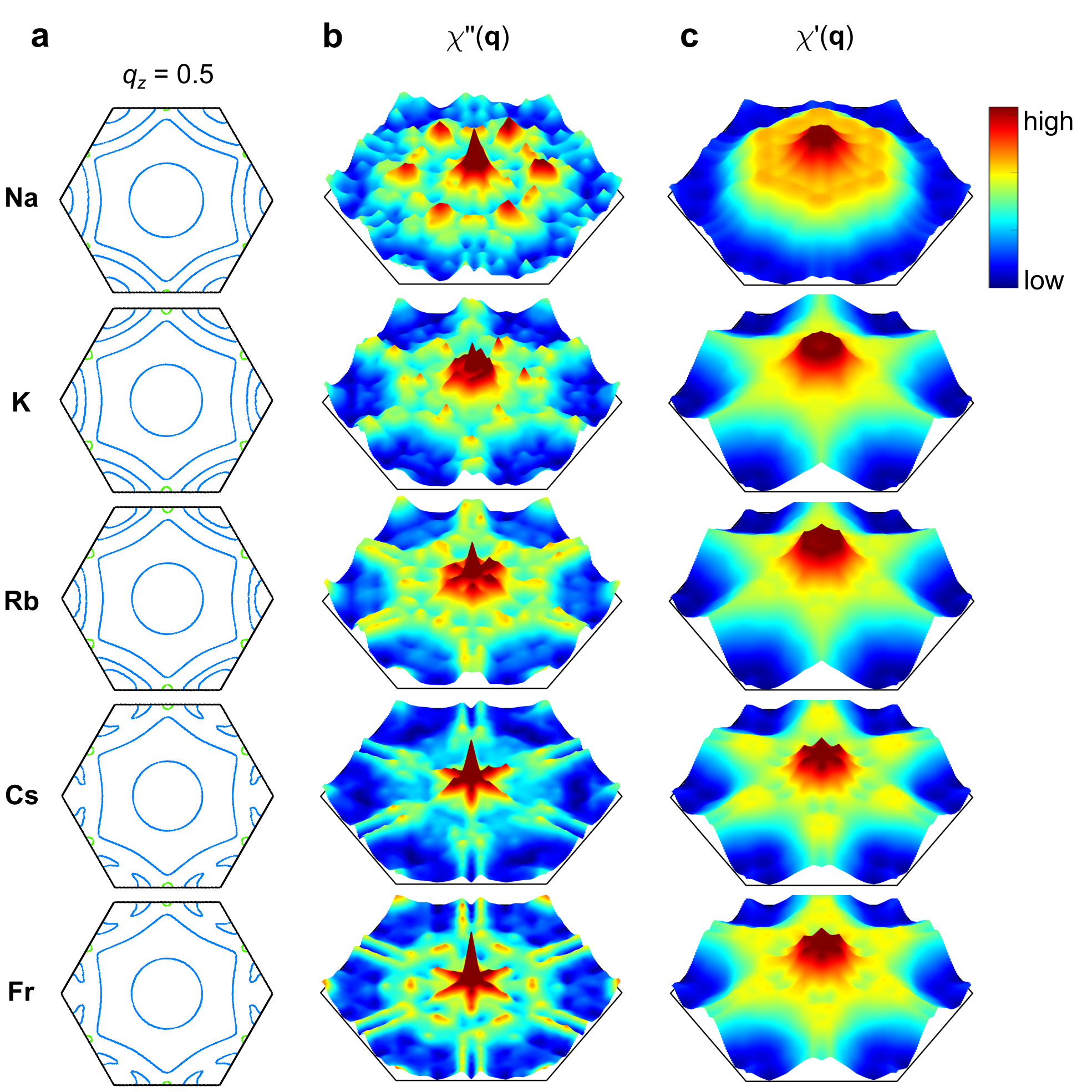}
	\caption{\textbf{a}~Fermi surfaces for different alkali ions at $q_z = 0.5$, illustrated using \texttt{FermiSurfer}~\cite{Kawamura2019}. All calculations were performed using experimental lattice parameters of \KRC. \textbf{b}~Imaginary part of the electronic susceptibility as a function of $q_x$ and $q_y$, with $q_z = 0.5$. \textbf{c}~Corresponding real part of the electronic susceptibility. }		
	\label{FS}
\end{figure}

In Fig.~\ref{FS}, we summarize changes in the Fermi surface at $q_z = 0.5$ as a function of $A$-site ion, based on calculations performed on \KRC. The results confirm the interpretation drawn from the $q_z = 0$ data presented in Fig.~4 of the main text. Although geometrical nesting is modified by the choice of alkali atom, absence of well-isolated, sharp peaks in $\chi'(\mathbf{q})$ indicates that the CDW instability is not associated with Fermi-surface nesting.

\begin{figure}[h]
	\centering
	\includegraphics[width=0.9\columnwidth]{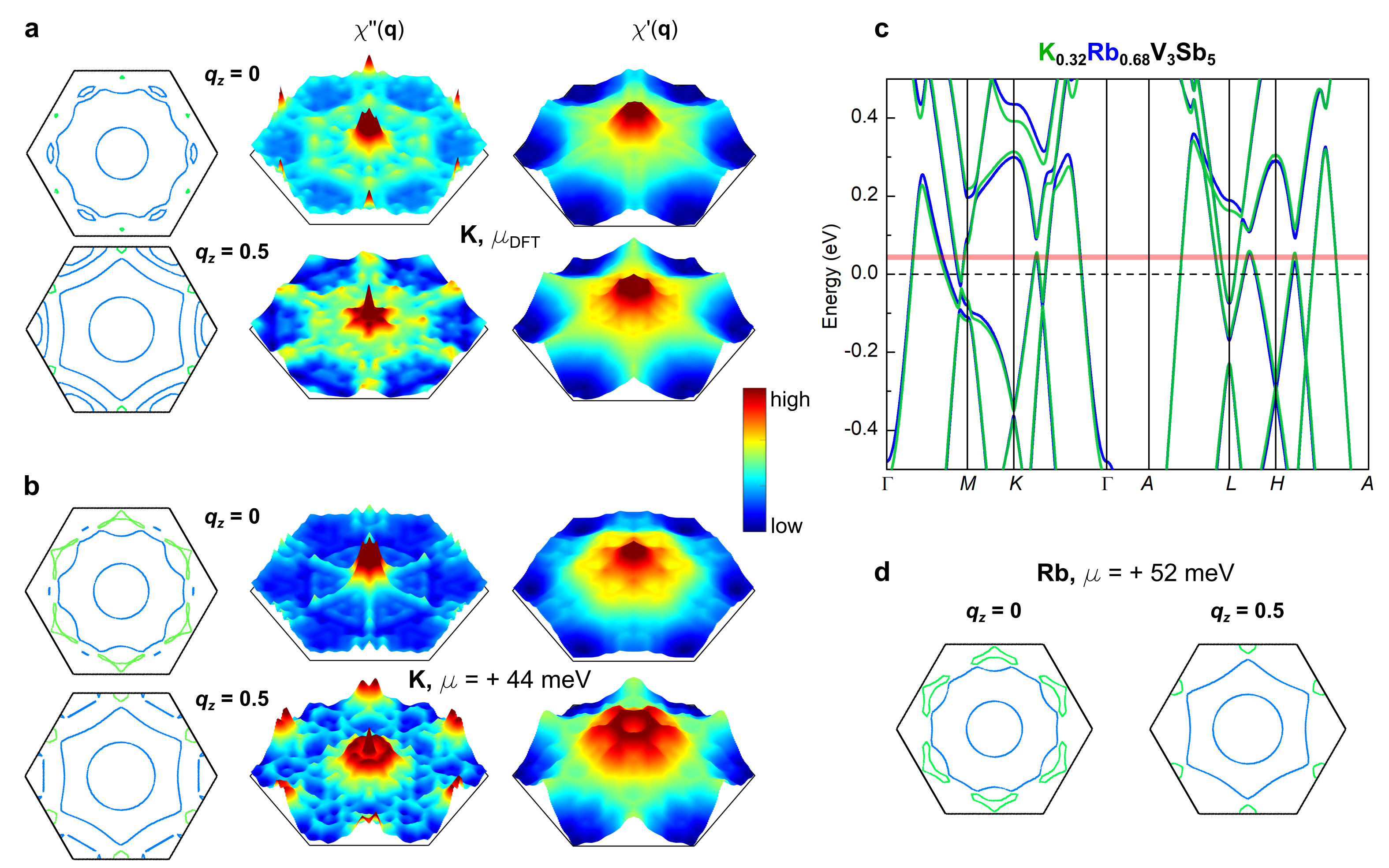}
	\caption{\textbf{a}~Fermi surface, imaginary, and real part of the electronic susceptibility as a function of $q_x$ and $q_y$, with $q_z = 0$ (upper row) and $q_z = 0.5$ (bottom row). The calculations were performed using K atoms and the lattice parameters of K$_{0.32}$Rb$_{0.68}$V$_3$Sb$_5$. The Fermi level is set to the DFT value. \textbf{b} Same but with the Fermi level adjusted using the experimental optical conductivity ($\mu = + 44$~meV). \textbf{c} Band structure of K$_{0.32}$Rb$_{0.68}$V$_3$Sb$_5$ calculated using K (green) or Rb atoms (blue). The adjusted Fermi level is indicated by the vertical red line. \textbf{d} Fermi surface of K$_{0.32}$Rb$_{0.68}$V$_3$Sb$_5$ calculated with Rb occupying the $A$-site and the Fermi level raised by + 52~meV at $q_z = 0$ (left) and $q_z = 0.5$ (right).}		
	\label{FS_shift}
\end{figure}

The optical spectra of the Cs-free compounds are reproduced only by introducing an upward shift of the Fermi level. In Fig.~\ref{FS_shift}, we present DFT calculations of the Fermi surface of K$_{0.32}$Rb$_{0.68}$V$_3$Sb$_5$. Although the moderate electron doping associated with the Fermi-level shift considerably modifies the Fermi surface and results in different geometrical nesting vectors, $\chi'(\mathbf{q})$ remains largely unaffected. This insensitivity of the electronic susceptibility to the modified Fermi-surface geometry further confirms the absence of an electronic instability associated with Fermi-surface nesting across all \AVS\ samples.

\subsection{Electronic Correlations}
The comparison between experimental and DFT-derived interband transitions reveals that Cs-containing \AVS\ samples are well described by band theory, whereas Cs-free compounds show prominent deviations that can be captured by a rigid upward shift of the Fermi level. Our previous optical spectroscopy studies have rendered \CVS\ essentially uncorrelated, while \RVS\ and \KVS\ show enhanced electronic correlations. Since, the kagome layer is only marginally affected by changes in the $A$-site, in-plane kagome interactions can be ruled out as the driving factor behind the evolution of the electronic correlations in \AVS\ compounds, leaving the significant changes in kagome interlayer distance (see Fig.~\ref{latticepara}) as a plausible driving factor behind these changes.

The comparison between the experimental and the DFT-derived (uncorrelated band theory) plasma frequency can be used to gauge the strength of electronic correlations by taking the ratio
\begin{equation}
\frac{\omega^2_{\mathrm{p, intra}}}{\omega^2_{\mathrm{p, DFT}}}.
\end{equation}
The experimental plasma frequency is determined from the intraband spectral weight (SW) analysis according to the sum rule \cite{Dressel2002}
\begin{equation}
\mathrm{SW} = \int_0^{\omega_{\mathrm{c}}} \sigma_{\mathrm{1}}(\omega) \mathrm{d}\omega = \frac{\omega_{\mathrm{p}}^2}{8}.
\end{equation}
The experimental intraband spectral weight is determined by integrating the real part of the optical conductivity of the fitted Drude and localization peak, choosing the cut-off frequency as $\omega_c = 50000$~\cm, considering the high-energy tail of the localization peak. This results in
\begin{equation}
\omega_{\mathrm{p, intra}} = \sqrt{8 \cdot (\mathrm{SW_{Drude}} + \mathrm{SW_{localization}})}.
\end{equation}

While the DFT-derived plasma frequencies are only marginally affected by the choice of alkali atom, we assume error bars of 10\%, accounting for uncertainties in determining the experimental spectral weight of the intraband response. The experimental plasma frequencies were determined at 100~K ($T > $~\Tc). The data highlight that a larger interlayer distance weakens electronic correlations, making pristine \CVS\ the least correlated compound in the \AVS\ family. Conversely, chemical substitution with smaller alkali atoms reduces interlayer spacing, which in turn increases the electronic correlation strength.

\begin{figure}[h]
	\centering
	\includegraphics[width=0.56\columnwidth]{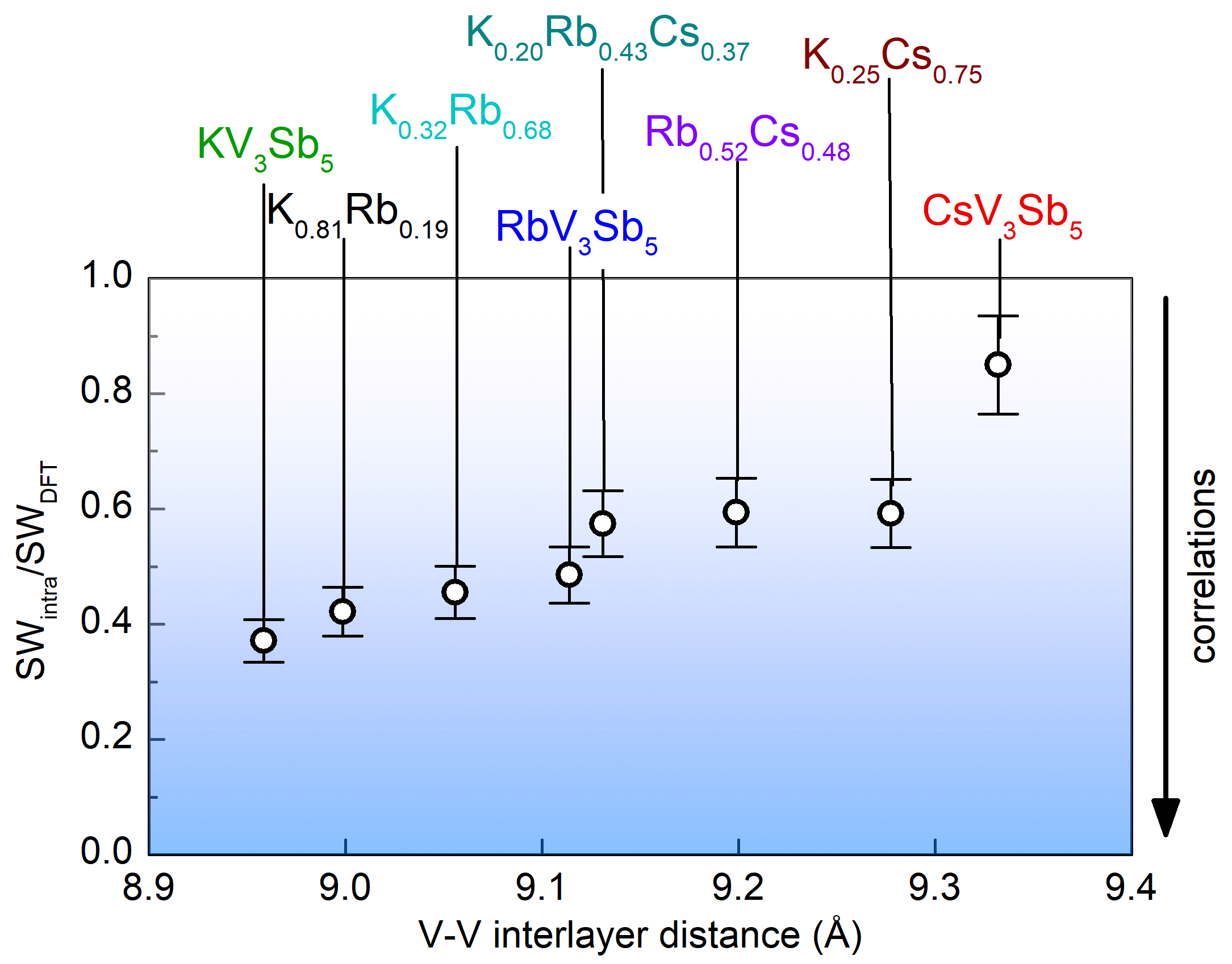}
	\caption{Ratio of the experimental and the DFT-based intraband spectral weights as a function of vanadium interlayer distance. Values for the parent compounds are taken from Refs.~\cite{Uykur2021, Uykur2022, Wenzel2022}. Error bars of 10\% are assumed due to uncertainties in determining the experimental intraband spectral weight.}		
	\label{correlations}
\end{figure}

\bibliography{mixedAsite.bib}